\documentclass[aps,prx,twocolumn,showpacs,preprintnumbers,amsmath,amssymb,superscriptaddress,longbibliography]{revtex4-2}

\usepackage{graphicx}% Include figure files
\usepackage{dcolumn}% Align table columns on decimal point
\usepackage{bm}% bold math
\usepackage{color}
\usepackage{ulem}
\usepackage{float}
\usepackage[colorlinks,bookmarks=true,citecolor=blue,linkcolor=blue,urlcolor=blue, breaklinks=true]{hyperref}

\begin{document}

\title{
Positronium resonance of electric monopoles in quantum spin ice
}% Force line breaks with \\

\author{Jos\'{e} M. Cruz}
%\affiliation{CeFEMA-LaPMET, Departamento de F\'{i}sica, Instituto Superior T\'{e}cnico, Universidade de Lisboa, Av. Rovisco Pais, 1049-001 Lisboa, Portugal}
\affiliation{Department of Physics, Gakushuin University, Mejiro, Tokyo, 171-8588, Japan}
\affiliation{School of Physics, University College Dublin, Dublin, D04V1W8, Ireland}
\email{jose.m.cruz@ucdconnect.ie}
\author{Hidemaro Suwa}
\affiliation{Department of Physics, The University of Tokyo, Tokyo 113-0033, Japan}
\author{Masafumi Udagawa}
\affiliation{Department of Physics, Gakushuin University, Mejiro, Tokyo, 171-8588, Japan}
\email{masafumi.udagawa@gakushuin.ac.jp}

\date{\today}% It is always \today, today,
             %  but any date may be explicitly specified

\begin{abstract}
Quantum spin ice attracts considerable attention due to its remarkable low-energy properties, described by the effective gauge theory, which mimics the quantum electrodynamics (QED). 
The ``effective" QED in quantum spin ice is beyond the ``real" QED in our universe in many respects. Above all, the effective QED hosts two kinds of gapped particles, the electric monopole, in addition to the magnetic monopole.
Among the two monopoles, the electric monopole is of purely quantum origin, and its character has not been well resolved.
In this work, we focus on the Rokhsar Kivelson point of the quantum dimer model, as the low-energy description of the quantum XXZ model on a pyrochlore lattice, and address the excitation spectrum of the electric monopoles, by combining the quantum Monte Carlo simulation with the stochastic analytical continuation.
Most strikingly, the electric monopole spectrum exhibits clear double-peak structure, corresponding to the bound states and scattering states of electric monopoles.
The lower-energy peak is attributed to the exciton-like bound state of an electric monopole pair, a signal of positronium in quantum spin ice.
We found that the flippable loop operator serves as a local physical observable to excite a pair of electric monopoles, implying that monopoles can be detected through lattice or optical response.
\end{abstract}

\maketitle
\section{Introduction}
The exotic properties of quantum spin liquid (QSL) continue to attract considerable attention in condensed matter physics~\cite{ANDERSON1973153,balents2010balents,annurev:/content/journals/10.1146/annurev-conmatphys-031218-013401,moessner2021topological}.
Among the surprising phenomena QSL exhibits, fractionalization deserves special attention~\cite{Lake:2010aa,Mourigal:2013aa,PhysRevLett.111.137205,de-Picciotto:1997aa,PhysRevLett.79.2526,doi:10.1126/science.aaz5601,Nakamura:2020aa}.
As a representative system of QSL, quantum spin ice (QSI) provides an interesting stage where fractionalization takes place in a unique way.

In QSI, single spin is fractionalized into a pair of magnetic monopoles.
In addition to this relatively simple route to fractionalization, another type of fractionalized particle, called an electric monopole, can be constructed out of the massive quantum coherence of the ground state~\cite{PhysRevB.69.064404}.
Among the two species of monopoles, magnetic monopoles exist even at the classical limit, which enables us to grasp them intuitively and makes them accessible experimentally in higher energy scale.
Whereas the absence of trivial limit makes it difficult to understand the electric monopole conceptually, and it is still outside experimental observability, either.
In this work, we focus on this electric monopole, and try to access its elusive characters.

To address the electric monopole, we start with a canonical model of spin ice.
Spin ice is a classical magnetic system defined on a pyrochlore lattice~\cite{Udagawa2021}, where the spins on each tetrahedron satisfies the 2-in 2-out ice rule~\cite{10.1063/1.1749327}, mimicking the configurations of hydrogen atoms in water ice~\cite{10.1021/ja01315a102}.
The ice rule leaves macroscopic degeneracy in the ground state, as observed through the residual entropy in Dy$_2$Ti$_2$O$_7$~\cite{Ramirez:1999aa}. 
Fractionalization already shows up in this classical level: the single spin flip from the ground state is fractionalized to a pair of magnetic monopoles with positive and negative magnetic charges~\cite{castelnovo2008magnetic}.
This ``magnetic monopole" is sometimes called ``spinon" or even ``electric monopole" in different terminology. 
In this work, we adopt ``magnetic monopole" for this particle, respecting its connection to the dipole magnetic moment.

QSI is the quantum superposition of the macroscopically degenerate classical spin ice configurations, which serves as a prototype of QSL~\cite{Gingras_2014}.
There are a macroscopic number of different ways of superposing 2-in 2-out spin ice configurations, and those superposed states orthogonal to the ground state make the low-energy excitations, involving the quasiparticles known as photons and electric monopoles.
The dynamics of these emergent quasiparticles is described by a familiar gauge theory, QED, mimicking the fundamental law in our real world~\cite{PhysRevB.86.075154}.
A number of materials have been proposed as candidates of quantum spin ice to host effective QED~\cite{Kimura:2013aa,Tang:2023aa,Uehara:2022aa,PhysRevB.94.165153,PhysRevLett.118.107206,PhysRevB.94.144415,Sibille:2018aa,Gao:2019aa,PhysRevLett.122.187201,Sibille:2020aa,Petit:2016aa,PhysRevLett.126.247201,tokiwa2016tokiwa,4qxy-l8pg,PhysRevB.111.155137,PhysRevB.106.094425,PhysRevX.12.021015,PhysRevB.108.054438,PhysRevX.15.021033,Gao:2025aa,Poree:2025aa}.
In fact, inelastic neutron scattering experiments on Ce$_2$Sn$_2$O$_7$ recently revealed the promising signatures for the quantum dynamics of magnetic monopoles~\cite{Poree:2025aa}, for which nontrivial features beyond conventional quasiparticles are expected~\cite{PhysRevB.96.085136,PhysRevLett.122.117201,PhysRevLett.124.097204}.

The QED realized in quantum spin ice has a lot in common with the QED in our real world, but it is, in fact, much beyond the real QED.
Recent numerical study of model quantum spin ice shows that the hyperfine constant of effective QED is close to $\alpha_{\rm eff}\sim0.1$~\cite{PhysRevLett.127.117205}, in contrast to the weak-coupling QED, $\alpha\sim1/137$ in our world.
It is even possible to tune the interaction attractive between same charge of monopoles~\cite{PhysRevB.94.104416,PhysRevLett.119.077207,rau2016spin}, as in Dyson's thought experiment~\cite{PhysRev.85.631}.
Concerning the deviation from the standard QED, the most interesting aspect may be the coexistence of two kinds of monopoles.
In our real world, the coexistence is (still) out of reach:
Magnetic monopoles have not been discovered, except for one possible observation~\cite{PhysRevLett.48.1378}, despite the intensive search even on the moon~\cite{doi:10.1126/science.167.3918.701}.
The situation in quantum spin ice is similar but opposite: the magnetic monopole is observed through, e.g. the noise in magnetic induction~\cite{Dusad:2019aa}, and it is the electric monopole that escapes experimental observation.

Search for electric monopoles may not be easy even theoretically.
After the theoretical proposal of the electric monopole, 
several intensive numerical analyses have been carried out for the microscopic models of QSI~\cite{Lauchli:2008aa,PhysRevLett.120.167202,PhysRevLett.100.047208}.
Even though they successfully identified low-energy excitations, e.g. photons, electric monopoles did not show up in their spectra.
Successively, important properties of electric monopoles have been clarified in various levels of analysis, such as the excitation gap~\cite{PhysRevB.95.134439,PhysRevB.100.014417}, momentum fragmentation~\cite{PhysRevB.96.195127} and crystallization~\cite{PhysRevB.109.094426,PhysRevB.73.134402}.
Provided the level of maturity of QSI study, it is desirable to address the basic properties of electric monopoles through a solid analysis of a microscopic model.

To this aim, we perform microscopic analysis by adopting the quantum dimer model as the model of quantum spin ice, and solve the model at the solvable Rokhsar-Kivelson point with rigorous quantum Monte Carlo simulation.
We further use the cutting-edge technique of stochastic analytical continuation (SAC) to transform the imaginary-time correlation function into the spectrum in the real-frequency representation.

The structure of the paper is as follows.
In Sec.~\ref{sec:ModelMethod}, we introduce the quantum dimer model and the variational wavefunction for a pair of electric monopoles at the Rokhsar-Kivelson point.
We also introduce the method to derive the spectral information from the imaginary-time correlation function, putting emphasis on the technique of the stochastic analytical continuation.
In Sec.~\ref{sec:Results}, we introduce our results. Starting with the low-energy photon excitations as a benchmark, we discuss the spectrum of electric monopoles, especially focusing on the single monopole energy gap and the interaction between monopoles. Our main result is the excitation spectrum excited by the flippable loop operator, which reveals the electric monopoles form exciton-like bound states, in addition to the scattering states.
Finally, in Sec.~\ref{sec:Summary}, we present summary and discussion.

\section{Model \& Method}
\label{sec:ModelMethod}
\subsection{Hamiltonian}
As a microscopic model to study the electric monopoles, we consider the quantum dimer model (QDM) on a diamond lattice.
First, to introduce this model, we start with the quantum XXZ model on a pyrochlore lattice,
\begin{eqnarray}
\mathcal{H} = J_z\sum_{\langle i,j\rangle}s^z_is^z_j - \frac{J_{\pm}}{2}\sum_{\langle i,j\rangle}(s^+_is^-_j+s^-_is^+_j),
\label{eq:XXZHamiltonian}
\end{eqnarray}
where $s^{\alpha}_j=\frac{1}{2}\sigma^{\alpha}_j$ is the spin-$1/2$ operator.
Here, $J_z$ is chosen antiferromagnetic, $J_z>0$, and $|J_{\pm}|\ll J_z$ is assumed.
If we completely ignore the exchange interaction, and set $J_{\pm}=0$, the ground state is macroscopically degenerate, composed of any spin configurations satisfying the ice rule:
on each tetrahedron, two spins satisfy $\sigma^z_j = +1$, and the other two satisfy $\sigma^z_j = -1$.

As another representation of this spin ice ground state, we introduce a dimer representation.
For this purpose, we first construct a diamond lattice, by placing a site at the center of each tetrahedron.
Then, by connecting neighboring sites with bonds, we obtain a diamond lattice. 
It means each vertex of pyrochlore lattice $j$ is connected to two sites on a diamond lattice.
We place a dimer on these two sites on the diamond lattice, if $\sigma^z_j = +1$ on the vertex $j$.
Consequently, ice rule configurations have one-to-one correspondence with the dimer representation on the diamond lattice, where each site on the diamond lattice is covered by exactly two dimers.

Within the limited Hilbert space composed of the 2-in 2-out spin ice configurations, or the equivalent dimer configurations, we define the Hamiltonian of the quantum dimer model~\cite{PhysRevB.68.184512},
\begin{eqnarray}
\mathcal{H} = -T + vV,
\label{eq:QDMHamiltonian}
\end{eqnarray}
where both $T$ and $V$ act on hexagonal plaquettes of the diamond lattice, which also corresponds to the hexagonal loop of pyrochlore lattice.
If dimers are placed in a staggered way on the plaquette, we call this plaquette as flippable.
If $T$ is applied on a flippable plaquette, $T$ rotates dimers, while if the plaquette takes different dimer configurations, $T$ simply vanishes the state.
$V$, is the projection on the flippable state, and this term expresses the potential energy to give penalty $v$ to a flippable plaquette. 
If we set $v=0$, the Hamiltonian (\ref{eq:QDMHamiltonian}) gives the effective description of XXZ Hamiltonian (\ref{eq:XXZHamiltonian}), derived by the third-order perturbation theory in terms of $\frac{J_{\pm}}{J_z}$.

Both terms of the Hamiltonian (\ref{eq:QDMHamiltonian}) have the representation of original spin degrees of freedom.
We write these terms as
\begin{eqnarray}
T = \sum_p (T_{p+} + T_{p-}), \ \ V = \sum_{\rm p}V_p,
\end{eqnarray}  
where the summation $\sum_{p}$ is taken over hexagonal loops. 
Then, they can be written by Pauli operators, as
\begin{eqnarray}
T_{p+} = \sigma^{+}_{p_1}\sigma^{-}_{p_2}\sigma^{+}_{p_3}\sigma^{-}_{p_4}\sigma^{+}_{p_5}\sigma^{-}_{p_6},\ \  T_{p-} = T^{\dag}_{p+},
\label{eq:PauliRepresentation_T}
\end{eqnarray}  
and 
\begin{align}
V_p &= \bigl(\frac{1+\sigma^z_{p_1}}{2}\bigr)\bigl(\frac{1-\sigma^z_{p_2}}{2}\bigr)\bigl(\frac{1+\sigma^z_{p_3}}{2}\bigr)\nonumber\\
&\hspace{2cm}\times\bigl(\frac{1-\sigma^z_{p_4}}{2}\bigr)\bigl(\frac{1+\sigma^z_{p_5}}{2}\bigr)\bigl(\frac{1-\sigma^z_{p_6}}{2}\bigr)\nonumber\\
 &+ \bigl(\frac{1-\sigma^z_{p_1}}{2}\bigr)\bigl(\frac{1+\sigma^z_{p_2}}{2}\bigr)\bigl(\frac{1-\sigma^z_{p_3}}{2}\bigr)\nonumber\\
&\hspace{2cm}\times\bigl(\frac{1+\sigma^z_{p_4}}{2}\bigr)\bigl(\frac{1-\sigma^z_{p_5}}{2}\bigr)\bigl(\frac{1+\sigma^z_{p_6}}{2}\bigr).
\label{eq:PauliRepresentation}
\end{align}  
Here, $p_1\cdots p_6$ are 6 sites of the hexagonal loop, $p$.

\subsection{Rokhsar-Kivelson point}
We focus on the Rokhsar-Kivelson (RK) point: $v=1$, where the ground state, $|\psi_{\rm RK}\rangle$, is expressed as the equal weight superposition of all possible spin ice configurations,
with the ground state energy, $E_0=0$.
At the RK point, we construct variational wavefunctions of excited states as
\begin{eqnarray}
|\psi\rangle\equiv\mathcal{O}|\psi_{\rm RK}\rangle, \hspace{1cm}\mathcal{O}\equiv\exp\Bigl[-\frac{i}{2}\sum_j\theta_j\sigma_j^z\Bigr],
\label{eq:variational_function}
\end{eqnarray}
by twisting spins at each site $j$, with different twisting angles, $\theta_j$.
With this operation, the operator $\mathcal{O}$ assigns different phases to the ice rule configurations in the macroscopic superposition, $|\psi_{\rm RK}\rangle$.

To evaluate the expectation values of energy, we consider the spin representation, Eqs.~(\ref{eq:PauliRepresentation_T}) and (\ref{eq:PauliRepresentation}).
Since $V$ commutes with $\mathcal{O}$, we obtain
\begin{align}
\langle\psi_{\rm RK}|\mathcal{O}^{\dag}V\mathcal{O}|\psi_{\rm RK}\rangle = \sum_{p}\langle\psi_{\rm RK}|V_{p}|\psi_{\rm RK}\rangle \equiv N_f,
\label{eq:Venergy}
\end{align}
where $N_f$ is the number of flippable plaquettes, and its ratio to the total number of plaquettes, $N_p$, is numerically evaluated as $\frac{N_f}{N_p}\sim0.175$~\cite{PhysRevB.69.064404}.

As for the ring exchange term, $T_p\equiv T_{p+} + T_{p-}$, we can evaluate its expectation value with the help of anticommutation relations of Pauli matrices, 
\begin{eqnarray}
\langle\psi_{\rm RK}|T_{p\pm}|\psi_{\rm RK}\rangle = \frac{N_f}{N_p}e^{\pm i\Phi_p},
\label{eq:RingexchangePhase}
\end{eqnarray}
where we introduced
\begin{eqnarray}
\Phi_p = \theta_{p1}-\theta_{p2}+\theta_{p3}-\theta_{p4}+\theta_{p5}-\theta_{p6}.
\end{eqnarray}
 Accordingly, we obtain the total energy
\begin{align}
\langle\psi_{\rm RK}|\mathcal{O}^{\dag}\mathcal{H}\mathcal{O}|\psi_{\rm RK}\rangle = \frac{N_f}{N_p}\sum_p(1-\cos\Phi_p).
\label{eq:VariationalEnergy}
\end{align}
For the derivation, see Appendix~\ref{sec:Evaluation_of_energy}.
By setting $\theta_j$ to be constant, we recover the ground state energy, $E_0=0$.

\subsection{Geometry}
\subsubsection{pyrochlore coordinates}
Before addressing electric monopole, we fix the convention of lattice geometries.
Spins (dimers) are defined on the vertices of pyrochlore lattice, $j$. The position of the vertex $j$ is determined by the unit cell index $i$ and the sublattice index, $\alpha$.
With these indices, we express $j=(i, \alpha)$, where $i=(i_1, i_2, i_3)$ and $\alpha=0, 1, 2, 3$. The position of the site is defined as
$\mathbf{R}_{i\alpha} = \mathbf{R}_{i} + {\mathbf e}_{\alpha}$,
where ${\mathbf e}_{\alpha}$ are 
\begin{eqnarray}
\hspace{-0.4cm}
{\mathbf e}_{0} = \begin{bmatrix}
0 \\ 0 \\ 0
\end{bmatrix}, 
{\mathbf e}_{1} = \begin{bmatrix}
0 \\ 1/2 \\ 1/2
\end{bmatrix}, 
{\mathbf e}_{2} = \begin{bmatrix}
1/2 \\ 0 \\ 1/2
\end{bmatrix}, 
{\mathbf e}_{3} = \begin{bmatrix}
1/2 \\ 1/2 \\ 0
\end{bmatrix},
\end{eqnarray}
in the $xyz$ coordinate. The position of unit cell is written as
\begin{eqnarray}
\mathbf{R}_{i} = i_1{\mathbf E}_1 + i_2{\mathbf E}_2 + i_3{\mathbf E}_3,
\end{eqnarray}
where ${\mathbf E}_{\alpha}=2{\mathbf e}_{\alpha}$ for $\alpha=0, 1, 2, 3$.

\subsubsection{dual pyrochlore lattice}
We here consider one supertetrahedron (of upward orientation), composed of four upward tetrahedra, where four hexagonal plaquettes make a closure.
We place a dual site at the center of each hexagonal plaquette, at which the electric fields will be defined later.
Then, the four dual sites defined on the four hexagonal plaquettes make a downward tetrahedron $i$, and the position of dual site or plaquette, is expressed as $p=(i, \alpha)$, whose coordinates we name as ${\mathbf r}_{i\alpha}$ ($\alpha=0,1,2,3$).
The unit cell index $i$ of the original pyrochlore lattice also serves as the unit cell index of dual lattice, and we define the origin of dual unit cell as
\begin{eqnarray}
{\mathbf r}_i = {\mathbf R}_i + ({\mathbf e}_0 +{\mathbf e}_1 + {\mathbf e}_2 + {\mathbf e}_3).
\end{eqnarray}
The coordinates of four sites on the dual downward tetrahedron are defined as
\begin{eqnarray}
{\mathbf r}_{i\alpha} = {\mathbf r}_{i} - {\mathbf e}_{\alpha}.
\end{eqnarray}
\begin{figure}[H]
    \centering
    \includegraphics[width=1\linewidth]{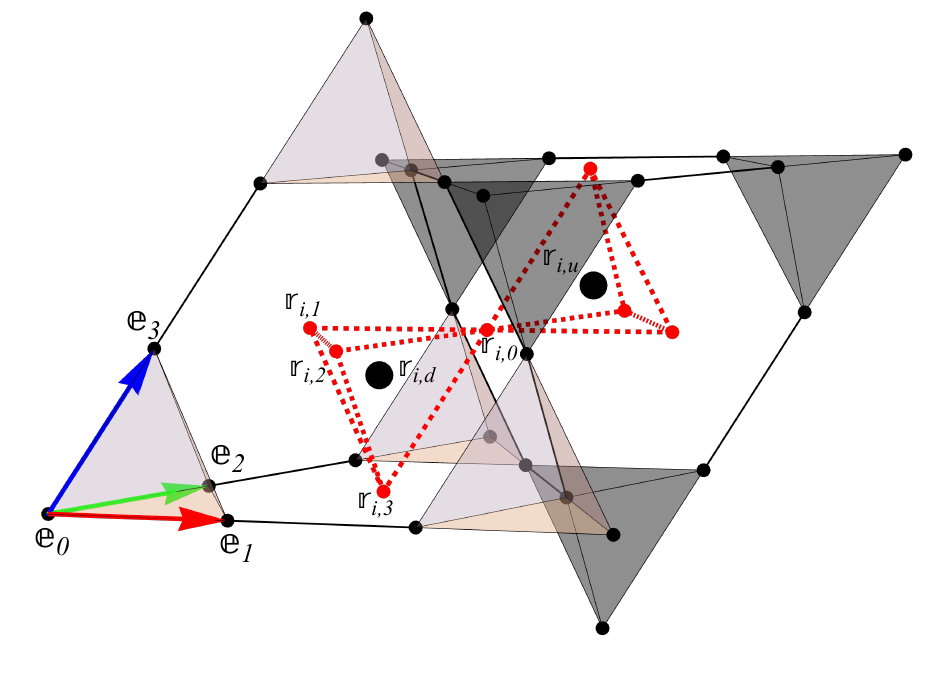}
    \caption{Downward and upward supertetrahedra, next to each other, of which the former (the latter) consists of four upward (downward) tetrahedra. a pair of downward and upward dual tetrahedra are shown with dashed red lines. The vertices of downward (upward) dual tetrahedron are located at the centers of hexagonal plaquettes of the upward (downward) supertetrahedron. The vertices of downward dual tetrahedron are denoted as ${\mathbf r}_{i\alpha}\equiv{\mathbf r}_{i}-{\mathbf e}_{\alpha}$, where the sublattice vectors ${\mathbf e}_{\alpha}$ ($\alpha=0,1,2,3$) are also shown. The electric fields $\Phi_{i\alpha}$ is defined at ${\mathbf r}_{i\alpha}$. The centers of the dual tetrahedra are denoted as ${\mathbf r}_{id}$ and ${\mathbf r}_{iu}$, at which electric monopoles are placed.}
    \label{fig:Supertetrahedron}
\end{figure}

It is convenient for later use to note that each site of the original pyrochlore lattice, ${\mathbf R}_{i\alpha}$ has 6 neighbors in the same original lattice at ${\mathbf R}_{i\alpha}\pm({\mathbf e}_{\alpha'} - {\mathbf e}_{\alpha'})$ for $\alpha'\not=\alpha$, and 6 neighbors in the dual lattice at ${\mathbf R}_{i\alpha}\pm({\mathbf e}_{\alpha'} - {\mathbf e}_{\alpha''})$ for $\alpha', \alpha''\not=\alpha$, and $\alpha'\not=\alpha''$.

Each unit cell $i$ contains two tetrahedra. 
We denote the coordinates of the center of downward and upward tetrahedra of the dual lattice as ${\mathbf r}_{id}$ and ${\mathbf r}_{iu}$, at which we electric monopole will be defined later.
Their positions are respectively expressed as
\begin{eqnarray}
\begin{cases}
{\mathbf r}_{id} = {\mathbf r}_{i} - \frac{1}{4}({\mathbf e}_0 +{\mathbf e}_1 + {\mathbf e}_2 + {\mathbf e}_3)\\
{\mathbf r}_{iu} = {\mathbf r}_{i} + \frac{1}{4}({\mathbf e}_0 +{\mathbf e}_1 + {\mathbf e}_2 + {\mathbf e}_3)
\end{cases}
\end{eqnarray}

\subsection{Electric monopole}
\label{sec:Emonopole}
A variational function for a pair of positive and negative electric monopoles can be constructed in a form, Eq.~(\ref{eq:variational_function}), by choosing a set of twisting angles $\{\theta_j\}$, properly~\cite{PhysRevB.69.064404}.
For the choice of $\{\theta_j\}$, it is helpful to make an analogy to classical electromagnetism, and regard $\{\theta_j\}$ as a vector potential defined at site $j$.
Then, we introduce a corresponding electric field $\Phi_{i\alpha}\sim\nabla\times\theta$ at each center of hexagonal plaquette on a pyrochlore lattice, or equivalently, a site of dual pyrochlore lattice, ${\mathbf r}_{i\alpha}$.
Then, the electric monopole charge is defined at a center of dual tetrahedron, ${\mathbf r}_{id}$ or ${\mathbf r}_{iu}$, as the divergence of the electric field $\Phi_{i\alpha}$.

As discussed in Eq.~(\ref{eq:RingexchangePhase}), $\Phi_{i\alpha}$ has the physical meaning of phase, associated with the ring exchange process.
If we virtually fractionalize a spin (dimer) into two spinons, and move one spinon around a hexagon, it results in the ring exchange process.
Accordingly, the electric field, $\Phi_{i\alpha}$, can be interpreted as Aharonov-Bohm flux for the fractionalized spinon.
We define the electric charge as the source of this Aharonov-Bohm flux.

To implement electric monopoles in the wavefunction, Eq.~(\ref{eq:variational_function}), we first solve Poisson's equation, i.e., we further impose the rotation-free condition for the electric fields, $\Phi_{i\alpha}$, and obtain the solution, $\{\theta_j\}$.
Then we incorporate this $\{\theta_j\}$ into Eq.~(\ref{eq:variational_function}), and obtain a variational wavefunction to describe the states with electric monopoles.

On the pyrochlore lattice, the procedure to solve Poisson equation is somewhat involved. The detailed derivation is summarized in Appendix~\ref{sec:Poisson}.
The final form of the electric field is given by
\begin{widetext}
\begin{eqnarray}
\Phi_{{\mathbf q},0}=\frac{|\vec{s}_{\mathbf q}|^2Q_{{\mathbf q},a} - i(\vec{s}_{\mathbf q}\cdot\vec{c}_{\mathbf q})Q_{{\mathbf q},s}}{|\vec{s}_{\mathbf q}|^2 + |\vec{s}_{\mathbf q}\times\vec{c}_{\mathbf q}|^2},
\label{eq:Phi0}
\end{eqnarray}
\begin{align}
\vec{\Phi}_{{\mathbf q}} = \frac{1}{|\vec{s}_{\mathbf q}|^2 + |\vec{s}_{\mathbf q}\times\vec{c}_{\mathbf q}|^2}\Bigl[\vec{s}_{\mathbf q}iQ_{{\mathbf q},s} + |\vec{s}_{\mathbf q}|^2\Bigl(\vec{c}_{\mathbf q} - \frac{\vec{s}_{\mathbf q}}{|\vec{s}_{\mathbf q}|}\Bigl(\vec{c}_{\mathbf q}\cdot\frac{\vec{s}_{\mathbf q}}{|\vec{s}_{\mathbf q}|}\Bigr)\Bigr)Q_{{\mathbf q},a} + |\vec{c}_{\mathbf q}|^2\Bigl(\vec{s}_{\mathbf q} - \frac{\vec{c}_{\mathbf q}}{|\vec{c}_{\mathbf q}|}\Bigl(\vec{s}_{\mathbf q}\cdot\frac{\vec{c}_{\mathbf q}}{|\vec{c}_{\mathbf q}|}\Bigr)\Bigr)iQ_{{\mathbf q},s}\Bigr].
\label{eq:Phi}
\end{align}
\end{widetext}
Here, we have introduced
\begin{eqnarray}
\vec{\Phi}_{{\mathbf q}} = \begin{bmatrix}
\Phi_{{\mathbf q},1}\\
\Phi_{{\mathbf q},2}\\
\Phi_{{\mathbf q},3}
\end{bmatrix},\ \ 
\vec{s}_{\mathbf q} = \begin{bmatrix}
\sin\frac{q_1}{2}\\
\sin\frac{q_2}{2}\\
\sin\frac{q_3}{2}
\end{bmatrix},\ \ 
\vec{c}_{\mathbf q} = \begin{bmatrix}
\cos\frac{q_1}{2}\\
\cos\frac{q_2}{2}\\
\cos\frac{q_3}{2}
\end{bmatrix}.
\end{eqnarray}

\begin{figure}
    \centering
    \includegraphics[width=1\linewidth]{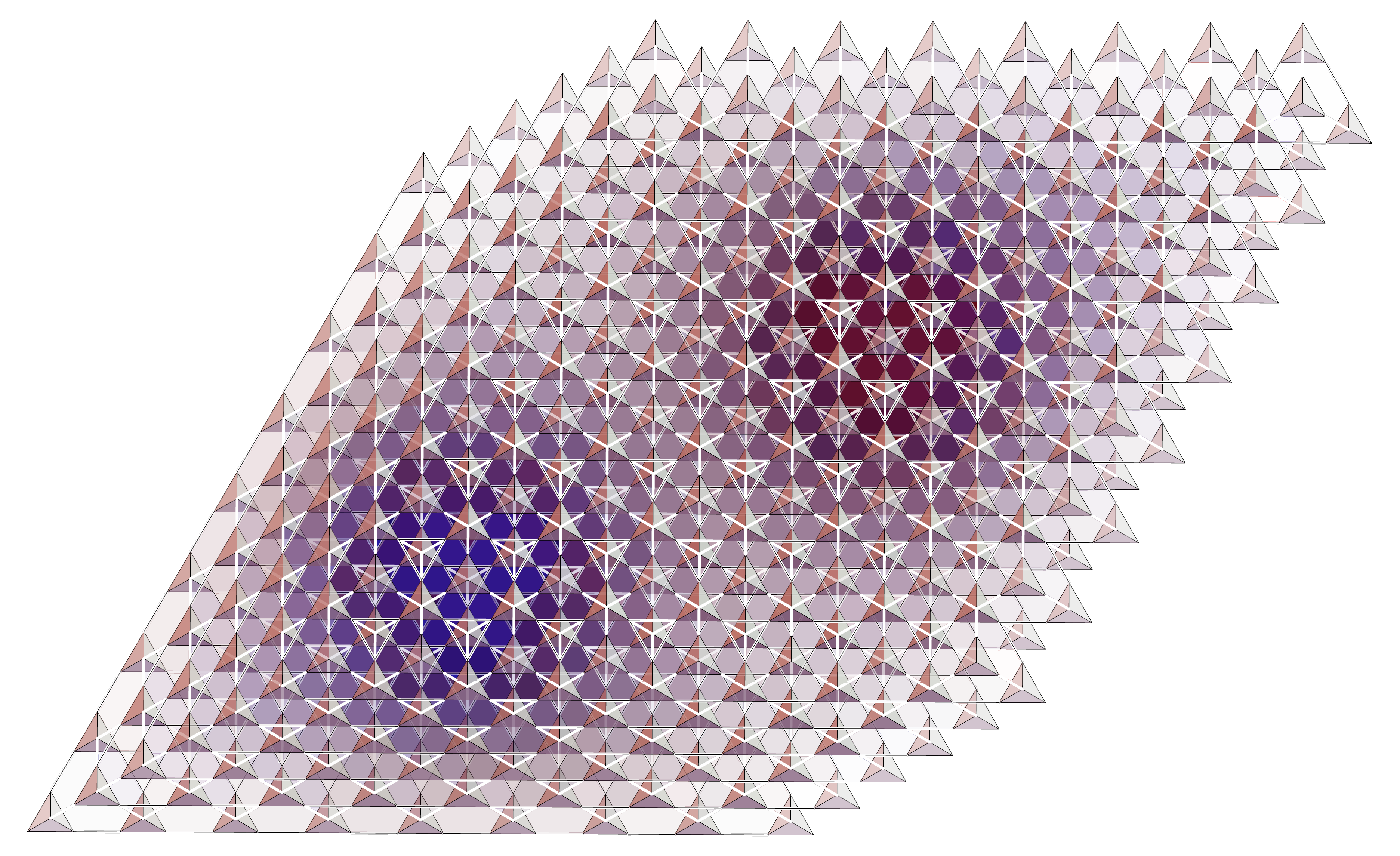}
    \caption{Electric fields are shown in color. In the system size of $L=10$, the positive and negative electric monopoles are placed along the line, ${\mathbf E_1}+{\mathbf E_2}+{\mathbf E_3}$. In the red (blue) region, the electric fields takes positive (negative) value, and their magnitudes are expressed by the strength of color.}
    \label{Fig:Fluxdistributions}
\end{figure}

\subsection{Method}
\subsubsection{Correlation function}
We consider the imaginary-time correlation function of operator $\mathcal{O}$, which is generally defined as
\begin{align}
G(\tau) &= \langle\mathcal{O}^{\dag}(\tau)\mathcal{O}(0)\rangle = \sum_n e^{-\tau(E_n - E_0)}|\langle n|\mathcal{O}|0\rangle|^2.
\end{align}
In the limit of infinite time, $\tau\to\infty$, $G(\tau)$ asymptotically becomes proportional to $e^{-\tau(E_1-E_0)}$.
Accordingly, in logarithmic scale, its slope gives first excitation gap, $E_1-E_0$.

From the technical viewpoint, the advantage to focus on RK point is that the correlation function $G(\tau)$ can be obtained from classical stochastic process~\cite{annurev:/content/journals/10.1146/annurev-conmatphys-070909-104138,CASTELNOVO2005316}.
We adopt kinetic Monte Carlo method to solve the stochastic equation numerically and obtain the imaginary-time correlation function.

\subsubsection{Stochastic analytical continuation}
 To access a broader energy landscape of the system beyond the first excitation energy, $E_1$, we resort to the stochastic analytical continuation, which has recently gained a lot of attention as a powerful method to obtain the spectral function~\cite{SHAO20231}.
We consider the zero-temperature spectral function $S(\omega)$ associated with an operator $\mathcal{O}$, given by
\begin{equation}
    S(\omega) = \pi\sum_{m} |\langle m|O|\psi_{\rm RK}\rangle|^2 \delta(\omega-E_m).
\end{equation}
The spectral function encodes rich information of the states excited by an operator, $\mathcal{O}$.

We obtain $S(\omega)$ from the imaginary-time correlation function, $G(\tau)$, which is obtained numerically by the kinetic Monte Carlo method, as they have the following relation

\begin{equation}
    G(\tau)=\frac{1}{\pi} \int^{\infty}_{-\infty} d\omega S(\omega)e^{-\omega\tau}
\end{equation}

We carry out this procedure through the stochastic analytical continuation (SAC), a Monte Carlo method which attempts to stochastically generate the spectral function by minimizing the ``goodness of fit" instead of the energy.
For the details of the procedure, see Appendix~\ref{appendix:SAC}.

\section{Results}
\label{sec:Results}
\subsection{Photon}
\begin{figure*}%[H]
   \centering
    \includegraphics[width=0.99\textwidth]{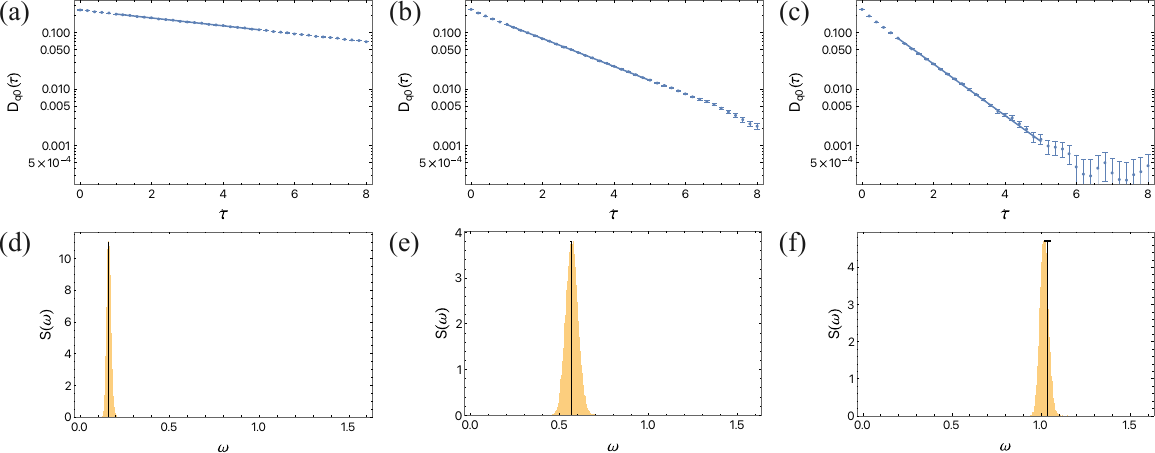} 
    \caption{Diagonal spin correlators $D_{{\mathbf q}\alpha}$ for $\alpha=0$, at (a) ${\mathbf q}=(\pi/6,\pi/6,0)$, (b) ${\mathbf q}=(\pi/3,\pi/3,0)$, and (c) ${\mathbf q}=(\pi/2,\pi/2,0)$. To estimate the excitation energy the slope was measured by fitting with result in the time range, $1<\tau<5$. 
    The corresponding spectral functions are shown in lower panels (d)-(f). The vertical lines in (d)-(f) are determined from the slopes of $D_{{\mathbf q}0}$ shown in (a)-(c). }
    \label{fig:SpinCorr}
\end{figure*}

We start with the lowest-energy excitation, photon, which is connected to an imaginary time correlation function of the spin operator, $D_{{\mathbf q}\alpha}(\tau)\equiv\langle\sigma^z_{\alpha}(-{\mathbf q}, \tau)\sigma^z_{\alpha}({\mathbf q}, 0)\rangle$.
Here, we focus on the diagonal component, in terms of the sublattice index.
 In Fig.~\ref{fig:SpinCorr}, we plot the time dependence of $D_{{\mathbf q}\alpha}(\tau)$ for several momenta.
 
Irrespective of the momenta, $D_{{\mathbf q}\alpha}(\tau)$ decays almost linearly from $\tau=0$ in the logarithmic scale. It means that $\sigma^z_{{\mathbf q}\alpha}|\psi_{\rm RK}\rangle$ is quite close to an eigenstate.
This state describes a single photon excitation. 
The excitation gap becomes small for smaller momenta, and closes at ${\mathbf q}=0$.
In particular, at Rokhsar Kivelson point, the gap closes quadratically $\propto|{\mathbf q}|^2$~\cite{Henley:2004aa,PhysRevB.68.184512}, which was recently proved as a rigorous theorem~\cite{PhysRevB.110.195140}. 

The full spectra of photons can be obtained by SAC [Fig~\ref{fig:SpinCorr} (d)-(f)]. 
For all the momenta, the spectrum consists of only a sharp single peak, suggesting that the spin operator excites a single particle with well-defined dispersion relation, consistent with the almost linear slope of $D_{{\mathbf q}\alpha}(\tau)$.
The position of the peak is located exactly at the energy obtained from the slope of $D_{{\mathbf q}\alpha}(\tau)$.
The agreement of two energies demonstrate that SAC indeed reproduces the excitation energy precisely.

\subsection{Electric monopole}
Now let us turn to electric monopole excitations.
As we discussed in Sec.~\ref{sec:Emonopole}, electric monopoles require a spatially extended representation in $\sigma_j^z$ basis, although we can regard monopoles as point-like excitations in the sense that they are topological defects of the electric fields centered at particular spatial points.

To construct a creation operator of an electric monopole pair, according to Eq.~(\ref{eq:variational_function}), 
we assume a particular monopole location; a positive monopole is at the downward tetrahedron at the origin, $(0,0,0)$, meanwhile a negative monopole is placed also at the downward tetrahedron of an arbitrary unit cell, $i=(i_1,i_2,i_3)$.
With this constraint, we solve the Poisson equation and obtain $\{\theta_j\}$, then construct an operator, $\mathcal{O}^{\rm E}_{i}=e^{-\frac{i}{2}\sum_j\theta_j\sigma^z_j}$.

We define a correlation function of an electric monopole pair by
\begin{eqnarray}
E_{i}(\tau)\equiv\langle\psi_{\rm RK}|\mathcal{O}^{{\rm E}\dag}_{i}(\tau)\mathcal{O}^{\rm E}_{i}|\psi_{\rm RK}\rangle.
\end{eqnarray}
From $E_{i}(\tau)$, we can extract the excitation gap to create an electric monopole pair, and also the entire excitation spectrum through the SAC.

\subsubsection{single monopole gap}
First, we discuss the excitation gap of an electric monopole pair.
We take the system of $L\times L\times L$ unit cells with $L=12$, and consider a spatial configuration that the positive monopole is located at $i=(l, 0, 0)$.
In other words, the two monopoles are separated along one of the lattice vectors, ${\mathbf E}_1$.

\begin{figure}[H]
\begin{center}
\includegraphics[width=0.5\textwidth]{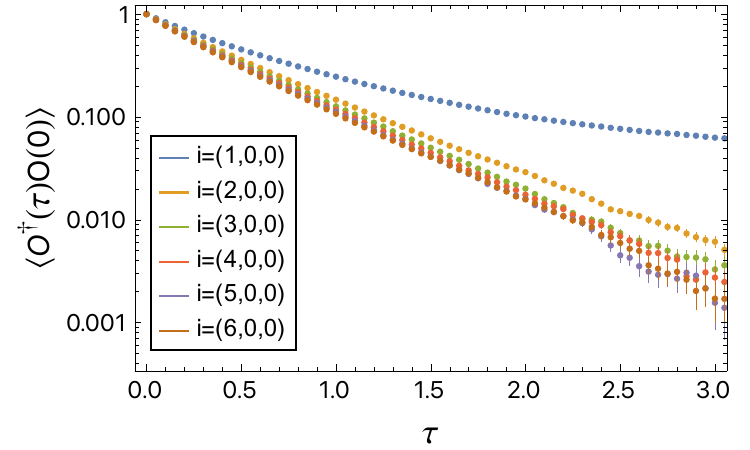}
\end{center}
\caption{\label{Fig:Emonopole_correlation1} 
(Color online) 
The time dependence of electric monopole correlation function, $E_{i}(\tau)$ for the configuration of $i=(L/2, 0, 0)$, in the system of $12\times12\times12$ unit cells ($L=12$).
}
\end{figure}

In Fig.~\ref{Fig:Emonopole_correlation1}, we plot the electric monopole correlation function, $E_{i}(\tau)$, with maximal separation length, $l=\frac{L}{2}$.
As is the case with the spin correlation function, $E_{i}(\tau)$ decays almost linearly in logarithmic scale, except for the case of $l=1$, i.e., the nearest-neighbor pair.
In the previous work~\cite{PhysRevB.69.064404}, it was reported that the the decay becomes nonlinear in a small system, as the overlap with the ground state cannot be negligible. 
In our system of $L=12$, this problem can be avoided for large separation.
However, if monopoles are placed in neighbor, they are subject to pair-annihilation to the ground state.

\begin{figure}[H]
\begin{center}
\includegraphics[width=0.5\textwidth]{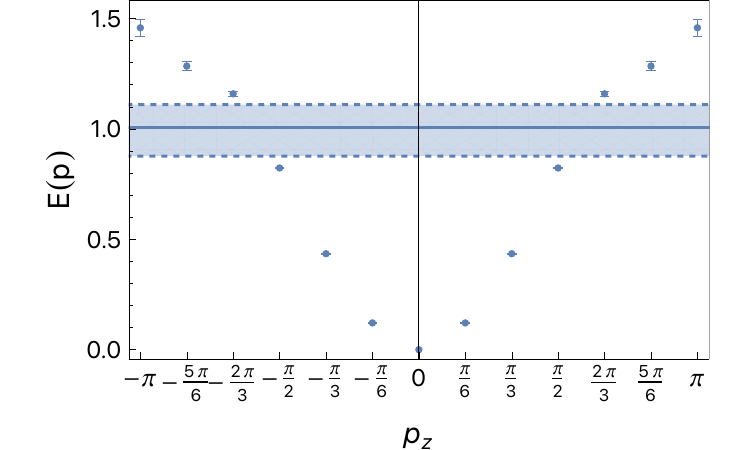}
\end{center}
\caption{\label{Fig:Emonopole_correlation2} 
(Color online) The solid line shows the mass gap $E_e$ of a single electric monopole, extracted from the slope of the correlation function, $E_{i}(\tau)$, at the maximal separation length, $i=(6, 0, 0)$. See Fig.~\ref{Fig:Emonopole_correlation1}. The photon dispersion is plotted with filled circles for comparison. The shade is the error bar of $E_{\rm e}$.}
\end{figure}

From the slope of the correlation function at the maximal separation, we estimate the excitation gap to be $2E_e\sim2.083$. 
Since $\mathcal{O}^{\rm E}_{i}$ creates two electric monopoles at the same time, the creation energy of single electric monopole is given by its half, $E_e=1.041$.
In Fig.~\ref{Fig:Emonopole_correlation2}, we compare $E_e$ with the photon dispersion.
Electric monopoles and photons have the same order of energy scale, determined by $J_{\rm ring}=\frac{3J_{\perp}^3}{2J_z^2}$, which we take to be 1.
Our analysis shows that, $E_e$ turns out to be slightly higher than the half of the photon band width at RK point.

It is interesting to compare the obtained single monopole gap, $E_e=1.041$, with the results of previous studies.
In the original work by Hermele et al.~\cite{PhysRevB.69.064404}, they reported an approximate estimate of $4.5J_{ring}$.
They introduced totally 4 electric monopoles, composed of 2 monopoles and 2 antimonopoles.
It means creation of one electric monopole costs $\sim 1.125J_{ring}$ on simple average, consistent with our estimate of $E_e=1.041J_{ring}$.

In the analysis at the semiclassical limit~\cite{PhysRevB.100.014417}, the excitation gap of $\mu=7.872367608$ is presented, in unit of $g=\frac{3J^3_{\perp}}{J^2}$. %Szabo
Although it seems much larger than our estimate, their theory led to a large photon band width, $\sim10g$, obtained from the dynamical structure factor.
Accordingly, the creation energy is $\sim0.7-0.8$ in unit of band width.
This ratio is quite close to our estimate above. 
Relative energy scale of single electric monopole and photon seems insensitive to the magnitude of spin.

\subsubsection{Coulomb interaction}
\begin{figure}[H]
\begin{center}
\includegraphics[width=0.5\textwidth]{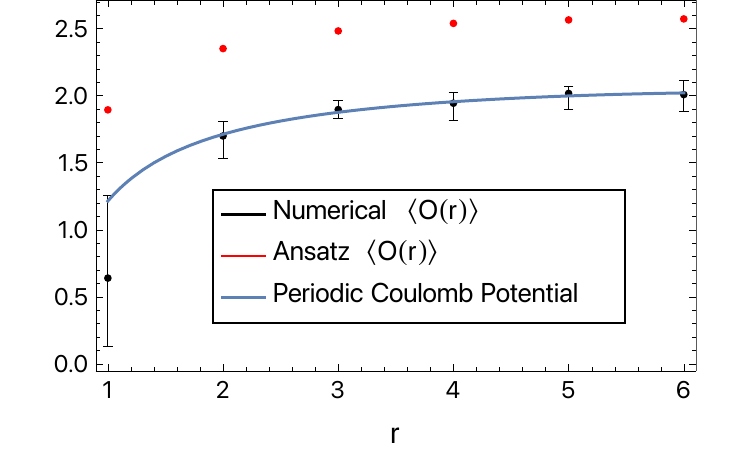}
\end{center}
\caption{\label{Fig:Emonopole_interaction} 
(Color online) The pair-monopole excitation gap is plotted against the separation, $l$. the blue solid curve shows the periodized Coulomb potential, $V(r)\propto\sum_n\frac{1}{|{\mathbf r} + {\mathbf n}L|}$. 
The red points are the variational energies.
}
\end{figure}

Next, we consider how the single monopole gap varies, depending on the separation between two monopoles, $l$.
The $l$-dependent excitation gap reflects the interaction energy between electric monopoles.
As plotted in Fig.~\ref{Fig:Emonopole_interaction}, the estimated excitation gap depends on the separation of monopoles, and it becomes smaller, as the monopoles are placed closer to each other, i.e. the two monopoles attract with each other.

The attractive interaction between the monopoles can be already seen in the form of variational parameters, Eqs.~(\ref{eq:Phi0}) and (\ref{eq:Phi}).
Roughly speaking, the energy cost of creating electric monopoles comes from the plaquette energy, $1-\cos\Phi_{p}$; if the electric field $\Phi_{p}$ takes a non-zero value, the plaquette $p$ contributes to the excitation energy.
Each monopole accompanies a cloud of high-energy plaquettes, as described in Fig.~\ref{Fig:Fluxdistributions}.
If two monopoles are placed closer to each other, the clouds overlap, and the volume of high-energy plaquettes is reduced.
From the form of variational wavefunctions, the interaction energy indeed becomes proportional to $-1/r$ in the long distance. See Appendix~\ref{sec:CoulombInteraction}.

In Fig.~\ref{Fig:Emonopole_interaction}, we compare the excitation energy of electric monopoles with its variational estimate, $\langle\psi_{\rm RK}|\mathcal{O}^{{\rm E}\dag}_{i}\mathcal{H}\mathcal{O}^{\rm E}_{i}|\psi_{\rm RK}\rangle$.
In the whole range of separation length, there is substantial relaxation from the variational energy.
As to the distance dependence, the expected $1/r$ behavior is not clear even for the variational energy in this system size, we can see similar tendency between the optimized and variational excitation energy; the latter approaches $1/r$ in the infinite system size. 

\subsubsection{Full spectrum}
\begin{figure}[H]
\begin{center}
\includegraphics[width=0.5\textwidth]{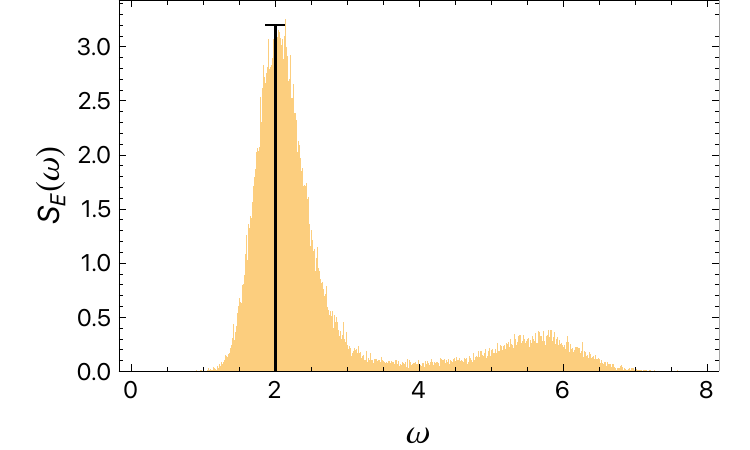}
\end{center}
\caption{\label{Fig:EmonopoleSpectrum} 
(Color online) Spectral function, $\mathcal{S}_{\rm E}(\omega)$, obtained from the imaginary-time electric monopole correlation function, $E_{i}(\tau)$ at the maximal separation, $i=(L/2,0,0)$.
Calculation is carried out for the system of $L\times L\times L$ unit cells with $L=12$. The vertical line is the excitation gap estimated from the slope of $E_{i}(\tau)$ in the logarithmic scale. }
\end{figure}

Now, we consider the full excitation spectrum of an electric monopole pair, $\mathcal{S}_{\rm E}(\omega)$.
For the analytical continuation, we use the correlation function, $E_{i}(\tau)$, obtained at the maximal separation, $i=(L/2, 0, 0)$.
The obtained spectrum is shown in Fig.~\ref{Fig:EmonopoleSpectrum}.
As a remarkable feature, the spectrum has two peaks.
In addition to the sharper low-energy peak around $\omega=2.0$, there is a second broad peak extended up to $\omega\sim7.0$.
This double-peak structure is the unprecedented discovery of this work, not reported in previous studies.

The position of the low-energy peak is compared with the excitation gap estimated from the slope of the electric monopole correlation function, $E_{i}(\tau)$. 
As is the case of the photon, the two estimates show good agreement.
The low-energy peak has somewhat larger width, compared with the photon peak at typical momenta; the height of this peak is reduced to half, meanwhile the half-peak width is $\sim4$ times larger.
As to the high-energy peak, it is difficult to find its signature in $E_{i}(\tau)$.
In addition to its small amplitude, the second peak is located at high energy, and decays quickly.
Without sophisticated technique like SAC used here, it would be difficult to reveal its existence.

\subsection{Positronum resonance}
While the electric monopole creation operator $\mathcal{O}^{\rm E}_i$ is suitable to access these excitations directly, the construction of the operator is nonlocal.
Now, we would like to ask how we can excite the electric monopoles with a local operator, desirably, connected to experimental observables.

\subsubsection{Flippable loop correlator}
To this aim, we define a flippable loop operator,
\begin{align}
&V_p = \prod_{j\in p}\frac{1+(-1)^{\eta_j}\sigma^z_j}{2} + \prod_{j\in p}\frac{1-(-1)^{\eta_j}\sigma^z_j}{2}.
\end{align}
$V_p$ is nothing but an operator constituting the potential term of the Hamiltonian, (\ref{eq:QDMHamiltonian}); $V_p$ returns $1$, if the hexagonal loop $p$ is flippable, and $0$, otherwise.
Coming back to the original spin Hamiltonian, (\ref{eq:XXZHamiltonian}), it means $V_p$ detects antiferromagnetic configuration around a loop, $p$.
We expect the antiferromagnetic spin correlation develops, if the bond length around a loop is shortened and the exchange interaction is enhanced.
Accordingly, we expect $V_p$ can be controlled by lattice distortion through e.g. ultrasonic attenuation and Raman scattering. 

\begin{figure}
    \centering   
     \includegraphics[width=1\linewidth]{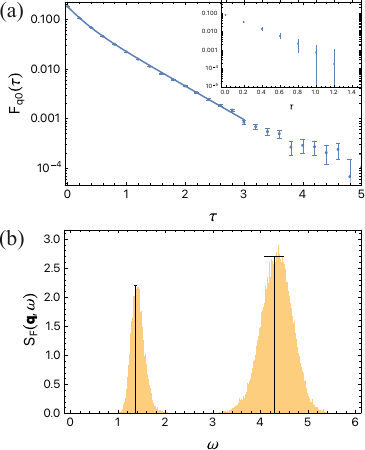}
    \caption{(a) Correlation function $F_{{\mathbf p}0}(\tau)$, evaluated at ${\mathbf q}=(\pi/2, \pi/2, \pi/2)$. The solid line shows the fitting curve by the two exponential ansatz, Eq.~(\ref{eq:twoexpfit}).
    The inset shows the remnant linear decay after subtracting the low-energy contribution.
    (b) The entire spectrum obtained from $F_{{\mathbf p}0}(\tau)$ at ${\mathbf q}=(\pi/2, \pi/2, \pi/2)$.
    }
    \label{Fig:tdepoffluxcorrelator}
\end{figure}

We move to momentum representation, $V_{{\mathbf q}\alpha}\equiv\sum_{i}e^{-i{\mathbf q}\cdot{\mathbf r}_{i\alpha}}V_{i\alpha}$, and define the correlation function,
\begin{eqnarray}
F_{{\mathbf q}\alpha}(\tau)\equiv\langle V_{-{\mathbf q},\alpha}(\tau)V_{{\mathbf q},\alpha}(0)\rangle.
\end{eqnarray}
The correlation function is plotted in Fig.~\ref{Fig:tdepoffluxcorrelator} (a) for ${\mathbf q}=(\pi/2, \pi/2, \pi/2)$ and $\alpha=0$.
As is the case with the spin correlation, $F_{{\mathbf q}0}(\tau)$ shows linear decay for large $\tau$ in logarithmic scale.
However, in contrast to the spin correlations, which show almost perfect linear decay at $\tau\sim0$, $F_{{\mathbf q}0}(\tau)$ shows slight bending at $\tau\sim0$, implying the non-negligible high-energy component is mixed in the state excited by the operator $V_{{\mathbf q}0}$ from the ground state.

We further make analytical continuation to obtain its full spectrum.
We show the spectrum for ${\mathbf q}=(\pi/2, \pi/2, \pi/2)$ in Fig.~\ref{Fig:tdepoffluxcorrelator} (b).
The spectrum consists of clear two peaks, where a sharp low-energy peak coexists with a broader high-energy peak.
The lower peak ranges from $\omega\sim1.0$ to $2.0$, while the higher peak ranges from $\omega\sim3.0$ to $5.5$.

To make the connection with the nonlinear decay of the correlation function displayed in Fig.~\ref{Fig:tdepoffluxcorrelator} (a), we fit the correlation function with two exponential functions,
\begin{eqnarray}
F_{{\mathbf q}0}(\tau) = A_{\mathbf q}e^{-E_L({\mathbf q})\tau} + B_{\mathbf q}e^{-E_H({\mathbf q})\tau},
\label{eq:twoexpfit}
\end{eqnarray}
where  $E_L({\mathbf q})<E_H({\mathbf q})$.

More specifically, we first estimate $E_L({\mathbf q})$ from the long time behavior of $F_{{\mathbf q}0}(\tau)$.
Then, we subtract the low energy part, $A_{\mathbf q}e^{-E_L({\mathbf q})t}$, and make linear fit to the remnant (inset of Fig.~\ref{Fig:tdepoffluxcorrelator} (a)), and estimate the high energy part, $E_H({\mathbf q})$.
By comparing $E_L({\mathbf q})$ and $E_H({\mathbf q})$ with the spectral peaks, as shown in Fig.~\ref{Fig:tdepoffluxcorrelator} (b), we find $E_L({\mathbf q})$ and $E_H({\mathbf q})$ respectively fit the position of low- and high-energy peaks.
The nonlinear decay of the correlation function means that the high-energy broader part of excitations decays quickly first, then the low-energy excitation shows slower decay, subsequently.

The double-peak structures of the electric monopole spectrum, $S_E(\omega)$ and the spectrum of the flippable loop operator, $S_F(\omega)$, suggest a common origin, although the weight of the peaks are different between them: In $S_F(\omega)$, both peaks have comparable heights, in contrast to those of $S_E(\omega)$.
We will come back to this point, later.
The common double-peak structure implies the possibility that the flippable loop operator excites electric monopoles, and therefore it serves as a useful probe for their detection.

To gain an insight into how the flippable loop operator $V_p$ couples to electric monopoles, let us consider the imaginary time evolution starting from the variation wavefunction $e^{-\frac{i}{2}\sum_j\theta_j\sigma^z_j}|\psi\rangle$.
Along with the imaginary time evolution, the system energy becomes gradually optimized, while keeping the electric monopoles unannihilated, at least for a short time.
As the formula of variational energy, Eq.~(\ref{eq:VariationalEnergy}) implies, the plaquettes with larger electric fields cost higher energy, $\propto 1-\cos\Phi_{p}$.
For more precise expression of energy in a more general state, see Appendix.~\ref{sec:Evaluation_of_energy}.
Consequently, the plaquettes with larger electric field, $\Phi_p$, becomes less flippable as a result of energy optimization through the imaginary time evolution.
It means that $V_p$ strongly couples to the line connecting a monopole pair, since $\Phi_p$ takes the maximal value between the neighboring electric monopoles.
This relation suggests $V_p$ creates a pair of electric monopoles on the both sides of the plaquette, $p$.
This speculation is supported by the position of the lower-energy peak in the spectrum, $S_F(\omega)$.
Its energy, $\omega\sim1.0-2.0$ is comparable to the excitation gap for the closest monopole pair, presented in Fig.~\ref{Fig:Emonopole_interaction}.

\subsubsection{Momentum resolution}
As already discussed, photon has vanishing excitation gap at ${\mathbf q}=0$.
In contrast, the excitations by the flippable loop operator have a finite gap at zero momenta.
The existence of the finite gap is reasonable.
In the case of photons, the long wave-length limit: $\theta_j={\rm Const.}$, corresponds to the uniform rotation of the spin, which does not generate a new state, and accordingly the excitation energy converges to zero.
Meanwhile, the ${\mathbf q}=0$ limit of the excitation by $V_{\mathbf q}$ physically modulates the state, and cost finite energy.

 \begin{figure}[H]
\begin{center}
\includegraphics[width=1\linewidth]{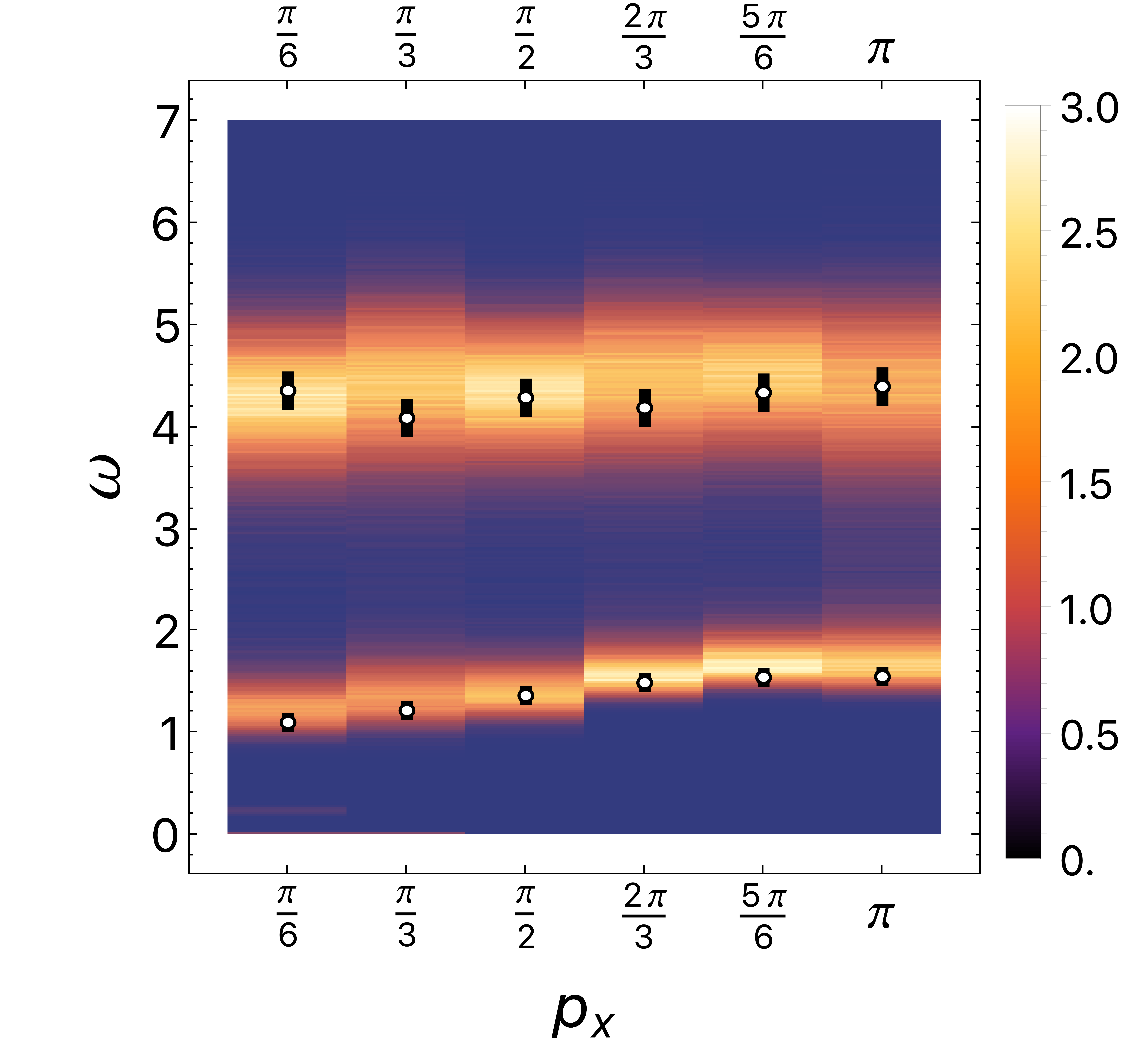}
\end{center}
\caption{\label{Fig:MomentumResolvedFluxSpectrum} 
(Color online) The momentum-energy plot of the spectrum, $S_F({\mathbf q}, \omega)$. The dots are the two energies, $E_{\rm L}({\mathbf q})$ and $E_{\rm H}({\mathbf q})$, obtained from the linear fit of the correlation function, $F_{\mathbf q}(\tau)$. 
}
\end{figure}

We plot the momentum-resolved spectrum in Fig.~\ref{Fig:MomentumResolvedFluxSpectrum}.
The spectrum consists of two energy scales, corresponding to the low- and high-energy peaks in Fig.~\ref{Fig:tdepoffluxcorrelator} (b).
Indeed, we again make the fits Eq.~(\ref{eq:twoexpfit}), and the obtained energies $E_{\rm L}({\mathbf q})$ and $E_{\rm H}({\mathbf q})$, which agree with the two energies of the spectra.
In the low energy scale, we can see the position of the peak depends on the momentum, and makes a slowly varying dispersion.
In contrast, the high energy part has smaller momentum dependence, and rather forms continuum structure.

The origin of the double-peak structure, obtained for the both spectra, $S_E({\mathbf q}, \omega)$ and $S_F({\mathbf q}, \omega)$, may be attributed to the two-particle states of an electric-monopole pair.
The low-energy sharp dispersion reflects the bound state of two electric monopoles.
Due to the Coulombic $1/r$ attractive potential, we expect two electric monopoles naturally form an excitonic bound pair, a positronium by electric monopoles.
This bound pair has finite center of mass, $M_{\rm e}=2m_{\rm e}$, and its motion makes a quadratic dispersion, $\varepsilon_{\mathbf q}\sim a + \frac{|{\mathbf q}|^2}{2M_{\rm e}}$.
The inertial mass of a single electric monopole can be estimated from the dispersion, to be $m_{\rm e}\sim2.14/J_{\rm ring}$. 

In contrast, the high energy continuum can be associated with the scattering state of the electric monopoles.
Again, the Coulombic $1/r$ attractive potential produces both bound state and scattering state.
This picture also accounts for the difference of relative amplitudes of low and high energy parts between electric monopole and flippable loop spectrum.
The electric monopole operator is constructed from the Poisson equation satisfied by the electric fields, $\Phi_i$.
It is analogous to the classical solution of static electric field, and accordingly, the corresponding charge has strong tendency of localization.
In contrast, the flippable loop operator is designed in a form of plane wave, which facilitates the coupling to scattering states with well defined momenta.

\section{Summary}
\label{sec:Summary}
To summarize, we have studied the dynamics of electric monopoles at the solvable RK point of the quantum dimer model, as an effective model of the quantum XXZ model on a pyrochlore lattice.
To access the information on dynamics, we used the quantum Monte Carlo simulation to obtain the imaginary-time correlation function.
We further used the stochastic analytical continuation to transform the imaginary-time correlation function into the spectral function in the real-frequency representation.

As a result, we found that the electric monopoles exhibit well-defined resonance peaks in their spectrum.
From the position of the peak, we estimated the single monopole gap to be $E_{\rm e}=1.041J_{\rm ring}$, which is consistent with the values obtained in the previous analyses~\cite{PhysRevB.69.064404,PhysRevB.100.014417}.
The position of the peak depends on the separation of monopoles, meaning the electric monopoles of the opposite signs of charge attract with each other.
Most strikingly, the spectrum shows clear double-peak structure, reflecting the scattering state and the bound state of a pair of electric monopoles.
In particular, the low-energy peak gives a signature of exciton-like bound state of electric monopoles, a counterpart of positronium in quantum spin ice.
From the dispersion of this exciton dispersion, we estimated the inertial mass of the single electric monopole to be $m_{\rm e}\sim2.14/J_{\rm ring}$.
The flippable loop operator shows similar double-peak spectra, implying this quantity, or more physically the lattice and optical responses are useful to detecting electric monopoles experimentally.

The present work can be extended to various directions.
Above all, our results give considerable insights into the gauge field dynamics in QSL in general.
Recently, gauge theories hosted by quantum spin ice was generalized to higher rank theories~\cite{PhysRevB.95.115139,PhysRevB.96.035119}, and the typical phase diagram has been established, taking account of the role of matter fields~\cite{6qmw-tktw}.
It is interesting to extend our theory to describe the dynamics of higher rank multipoles, which play a role of conserved charges in these theories.

In a class of QSL's in two dimensions, vison is known as a typical vortex-type gapped excitation, and its dynamics has been investigated~\cite{PhysRevB.70.094430,PhysRevB.74.134301}.
For example, dynamics of visons in Kitaev's spin liquids draws considerable attention~\cite{PhysRevB.98.220404,PhysRevX.12.041004,PhysRevB.107.045114,2r4t-tmqn}, due to the strong interest in the detection and manipulation of Majorana zero modes~\cite{PhysRevLett.126.127201,PhysRevLett.131.236701}.
There have also been long disputes over the QSL ground state of quantum Heisenberg model on kagome lattice and its candidate compounds, with renewed interest due to the discovery of $1/9$-th magnetization plateau in these systems~\cite{Nishimoto:2013aa,Okuma:2019aa,PhysRevB.93.060407,doi:10.7566/JPSJ.93.123706,Jeon:2024aa,PhysRevLett.132.226701,1vmv-zdqt}.
Search for visons may give a hint to identifying the QSL signature of the ground state.
We wish our results will be useful to the future study in these interesting problems.

\acknowledgments
 This work was supported by the JSPS KAKENHI (Nos. JP26K00634 and JP26K00662), MEXT, Japan.
 We thank Michael Hermele and Arnab Sen for their useful comments at the early stage of this study. 

\appendix
\section{Evaluation of energy}
\label{sec:Evaluation_of_energy}
We introduce the evaluation of energy expectation value.
Suppose a general form of wavefunction,
\begin{eqnarray}
|\psi\rangle = \sum_{\{\sigma\}}c_{\{\sigma\}}|\{\sigma\}\rangle,
\end{eqnarray}
where $|\{\sigma\}\rangle\equiv|\sigma_1^z\sigma_2^z\cdots\sigma_N^z\rangle$ is the spin state in the $\sigma^z$ basis, and the summation $\sum_{\{\sigma\}}$ is over all the $2^N$ spin configurations.
Taking the expectation value of Hamiltonian, we obtain
\begin{widetext}
\begin{eqnarray}
\langle\psi|\mathcal{H}|\psi\rangle = \sum_p\sum_{\{\sigma\}}\delta_p(\{\sigma\})\Bigl(\frac{|c_{\{\sigma\}}|^2 + |c_{\{\sigma_p\}}|^2}{2} - |c_{\{\sigma\}}||c_{\{\sigma_p\}}|\cos\Phi_p(\{\sigma\})\Bigr).
\end{eqnarray}
\end{widetext}
Here, $\delta_p(\{\sigma\})=1$, if and only if the plaquette $p$ is flippable in the spin configuration, $\{\sigma\}$, and $=0$, otherwise. $\{\sigma_p\}$ is the spin configuration obtained from $\{\sigma\}$, by flipping spins on the plaquette, $p$. $\Phi_p(\{\sigma\})$ is the phase, defined by $c^*_{\{\sigma\}}c_{\{\sigma_p\}}=|c_{\{\sigma\}}||c_{\{\sigma_p\}}|e^{i\Phi_p(\{\sigma\})}$.
From this expression, we find the weight $|c_{\{\sigma\}}|$ should be smaller for flippable plaquettes with large phase $\Phi_p(\{\sigma\})$, to lower the energy.

For the special type of variational wavefunction introduced in Eq.~(\ref{eq:variational_function}) in the main text, we can go further.
For the potential term, $V$, since it commutes with the operator, $\mathcal{O}=\exp\Bigl[-\frac{i}{2}\sum_j\theta_j\sigma_j^z\Bigr]$, its expectation value can be trivially evaluated as
\begin{align}
\langle\psi_{\rm RK}|\mathcal{O}^{\dag}V\mathcal{O}|\psi_{\rm RK}\rangle = \langle\psi_{\rm RK}|V|\psi_{\rm RK}\rangle = \frac{N_f}{N_p}.
\end{align}
It is equal to the ratio of the flippable plaquettes.
To evaluate the kinetic term, $T=\sum_pT_{p+} + T_{p-}$, we express them with Pauli matrices, then
\begin{widetext}
 \begin{align}
&\langle\psi_{\rm RK}|\mathcal{O}^{\dag}T_{p+}\mathcal{O}|\psi_{\rm RK}\rangle\nonumber\\
&= \langle\psi_{\rm RK}|e^{\frac{i}{2}(\theta_{p_1}\sigma_{p_1}^z+\theta_{p_2}\sigma_{p_2}^z+\theta_{p_3}\sigma_{p_3}^z+\theta_{p_4}\sigma_{p_4}^z+\theta_{p_5}\sigma_{p_5}^z+\theta_{p_6}\sigma_{p_6}^z)}\sigma^{+}_{p_1}\sigma^{-}_{p_2}\sigma^{+}_{p_3}\sigma^{-}_{p_4}\sigma^{+}_{p_5}\sigma^{-}_{p_6}e^{-\frac{i}{2}(\theta_{p_1}\sigma_{p_1}^z+\theta_{p_2}\sigma_{p_2}^z+\theta_{p_3}\sigma_{p_3}^z+\theta_{p_4}\sigma_{p_4}^z+\theta_{p_5}\sigma_{p_5}^z+\theta_{p_6}\sigma_{p_6}^z)}|\psi_{\rm RK}\rangle\nonumber\\
&= \langle\psi_{\rm RK}|e^{i(\theta_{p_1}\sigma_{p_1}^z+\theta_{p_2}\sigma_{p_2}^z+\theta_{p_3}\sigma_{p_3}^z+\theta_{p_4}\sigma_{p_4}^z+\theta_{p_5}\sigma_{p_5}^z+\theta_{p_6}\sigma_{p_6}^z)}\sigma^{+}_{p_1}\sigma^{-}_{p_2}\sigma^{+}_{p_3}\sigma^{-}_{p_4}\sigma^{+}_{p_5}\sigma^{-}_{p_6}|\psi_{\rm RK}\rangle\nonumber\\
&= e^{i(\theta_{p_1}-\theta_{p_2}+\theta_{p_3}-\theta_{p_4}+\theta_{p_5}-\theta_{p_6})}\langle\psi_{\rm RK}|\sigma^{+}_{p_1}\sigma^{-}_{p_2}\sigma^{+}_{p_3}\sigma^{-}_{p_4}\sigma^{+}_{p_5}\sigma^{-}_{p_6}|\psi_{\rm RK}\rangle
\label{eq:Tsquare_calc}
\end{align}
\end{widetext}

Treating the terms with $T_{p-}$ in a similar way, we obtain
\begin{widetext}
 \begin{align}
\langle\psi_{\rm RK}|\mathcal{O}^{\dag}T\mathcal{O}|\psi_{\rm RK}\rangle = \langle\cos(\theta_{p_1}-\theta_{p_2}+\theta_{p_3}-\theta_{p_4}+\theta_{p_5}-\theta_{p_6})\rangle_f,
\end{align}
\end{widetext}
where $\langle\cdots\rangle_f$ means the average over flippable plaquettes.
Combining with the contribution from $V$, we obtain
 \begin{align}
\frac{E}{N_f} = 1 - \langle\cos(\theta_{p_1}-\theta_{p_2}+\theta_{p_3}-\theta_{p_4}+\theta_{p_5}-\theta_{p_6})\rangle_f.
\end{align}

\section{Stochastic analytical continuation}
\label{appendix:SAC}
In this section, we introduce the technical points of stochastic analytical continuation (SAC).
SAC is the method to obtain the spectral function from the numerical data of imaginary time correlation function, through the relation,
\begin{equation}
    G(\tau) = \frac{1}{\pi} \int_{-\infty}^{\infty} d\omega S(\omega) e^{-\tau \omega},
\end{equation}
which is valid at zero temperature.
The data of $G(\tau)$ are numerically obtained, and inevitably contains numerical error.
In SAC, we stochastically deal with the numerical noise and constructs a spectrum $S(\omega)$ compatible with $G(\tau)$.

\begin{figure}[H]
    \centering
    \includegraphics[width=1\linewidth]{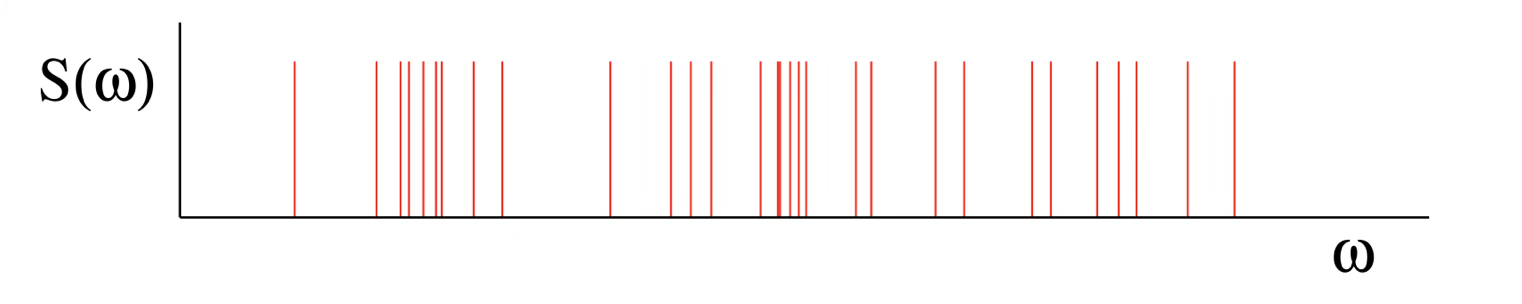}
    \caption{Example of initial guess for the spectrum $S_{\rm t}(\omega)$. Figure adapted from \cite{SHAO20231}.}
    \label{fig:InitialSpectrum}
\end{figure}

Specifically, we make Monte Carlo sampling of trial spectral functions.
As an initial guess for the spectrum, we assume a collection of randomly generated delta peaks, all with the same weight, as shown in Fig.~\ref{fig:InitialSpectrum},
\begin{equation}
S_{\rm t}(\omega) = \frac{1}{N_{\omega}}\sum_{k=1}^{N_\omega}\delta(\omega - \omega_{k}).
\label{eq:trialS}
\end{equation}
We set the number of peaks as $N_\omega=500$.
Then, we make the Brownian motion of the position of each delta peak.
We make a trial to choose peaks and move them to different positions, and decide if this trial is accepted or rejected, according to the assumed probability distribution.

In SAC, we assume the probability distribution to be the Gaussian form, $P=\exp(-\frac{\chi^2}{2\Theta})$.
To introduce this probability distribution, we discretize the imaginary time into $N_\tau$ equal intervals, and obtain the averaged Green function $\Bar{G}(i)$, at each time point, $t_i$.
$\Bar{G}(i)$ is obtained by taking sample average of Green function $G(i)$, obtained by the kinetic Monte Carlo method, over $N_s$ samples.
For most our simulation we used $N_\tau=25$ with a time step of $\Delta \tau = 0.2$ in normalized time units. 
We also generate trial value of Green function $G_{\rm t}(i)$ for the time $t_i$, by converting the trial spectral function $S_{\rm t}$ given in Eq.~(\ref{eq:trialS}).
Then, we define the ``goodness of fit", $\chi^2$ as
\begin{equation}
    \chi^2 = \sum_{i,j}^{N_{\tau}} (G_{\rm t}(i) - \Bar{G}(i))C_{ij}^{-1}(G_{\rm t}(j) - \Bar{G}(j)).
\end{equation}
Here, the covariance matrix, $C_{ij}$ contains the information of numerical error through the calculation of kinetic Monte Carlo simulation, which is defined as
\begin{equation}
    C_{ij} =\frac{1}{N_b(N_b -1 )}\sum_{b=1}^{N_b} (G^b(i) - \Bar{G}(i))(G^b(j) - \Bar{G}(j)).
\end{equation}
$N_b$ is the number of bins; we first divide $N_s$ samples of $G(i)$ into $N_b$ bins, each involving $N_s/N_b$ samples. 
And within each bin, we average $N_s/N_b$ samples to obtain $G^b(i)$. $\Bar{G}(i)$ is the averaged Green function as we noted above, which is of course equal to the average of  $G^b(i)$.

The main crux of this method is using this ``goodness of fit" $\chi^2$ as a replacement for energy in a Monte-Carlo algorithm, being this the quantity that we want to minimize. 
Every time we update the trial spectral function $S_{\rm t}$, The variation $\Delta \chi^2$ between the old configuration and this new configuration is calculated, and we accept the new configuration, according to the probability, $p=e^{-\frac{\Delta \chi^2}{2\Theta}}$, where $\Theta$ is a fictitious temperature.

The fictitious temperature $\Theta$ plays a crucial role to regulate the acceptance ratio of the updates. 
If it is very large, the spectrum will be too broad to make a satisfactory resolution.
If it is very small, most updates will not be accepted and we might run into convergence problems. 
We choose appropriate value of $\Theta$ by a simulated annealing procedure, following the strategy introduced in Ref.~\cite{SHAO20231}.
We start from high values of $\Theta$ and let the system thermalize and reach equilibrium. We confirm that equilibrium is reached after $10^6$ iterations. One iteration consists of attempting single, double and triple frequency moves:

\begin{itemize}
    \item \textbf{Single frequency move:} attempt to change $\omega_i \rightarrow \omega_i + R$, where $R$ is sampled randomly from $R\in [-50\sqrt{\Theta},50\sqrt{\Theta}]$. Having a temperature dependent interval is useful for convergence since we follow a annealing strategy. With lower $\Theta$, the spectrum will have converged more to the actual spectrum, so large frequency moves are more likely to be discarded. 
    \item \textbf{Double frequency moves:} Attempt to update two frequencies simultaneously $\omega_i \rightarrow \omega_i + R$ and $\omega_j \rightarrow \omega_j - R$, where R is chosen randomly from the same interval as the single frequency move. 
    \item \textbf{Triple frequency moves:} Attempt to update three frequencies simultaneously $\omega_i \rightarrow \omega_i + 2R$, $\omega_j \rightarrow \omega_j - R$ and $\omega_k \rightarrow \omega_k - R$, with $R=\frac{1}{3}(w_k+w_j-2w_i)$.
\end{itemize}

The advantage of the double and triple frequency moves is that they conserve higher-momenta of the spectrum, allowing for better fine-tuning while keeping broader features intact.

\begin{figure}[H]
    \centering
    \includegraphics[width=0.9\linewidth]{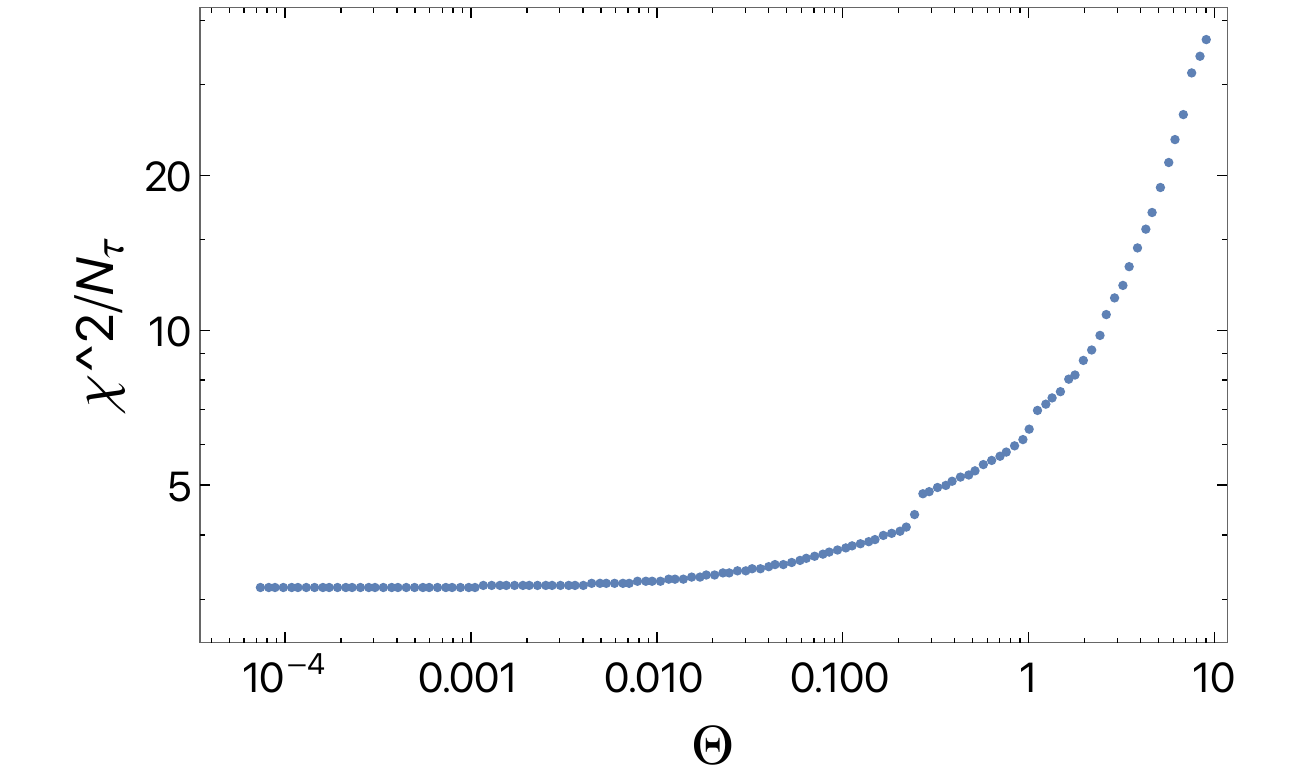}
    \caption{Annealing procedure for the spin correlation function for $q=(\pi/2,\pi/2,0)$. Each point corresponds of an average of 1000 $\chi^2/N_{\tau} $ samples for each fictitious temperature $\Theta$. From this process, we adopt $\Theta=0.001$ for the probability distribution, $P=\exp(-\frac{\chi^2}{2\Theta})$.}
    \label{fig:Annealing}
\end{figure}

We then repeat this procedure while slightly decreasing the value of $\Theta$ in each step until $\chi^2$ shows saturation. 
Then, we adopt the value of $\Theta$, just before $\chi^2$ shows the saturating behavior.
In the example shown in Fig.\ref{fig:Annealing}, for the spin correlation function, we choose $\Theta=0.001$.
This is the general strategy, and for each calculation in the main text, we confirmed the result is not sensitive to the slight change of $\Theta$.

\section{Poisson equation}
\label{sec:Poisson}
In this section, we introduce the procedure to determine $\{\theta_j\}$ by solving discretized Poission's equation, given the fixed distribution of electric monopoles.
First, we fix $Q_{id}$ and $Q_{iu}$, which respectively expresses the values of electric charges defined on the downward and upward tetrahedron at the unit cell, $i$.
They are connected with the electric fields through the relations,
\begin{eqnarray}
\begin{cases}
Q_{id} = \Phi_{{\mathbf r}_i} + \Phi_{{\mathbf r}_i-{\mathbf e}_1} + \Phi_{{\mathbf r}_i-{\mathbf e}_2} + \Phi_{{\mathbf r}_i-{\mathbf e}_3}, \\
Q_{iu} = -(\Phi_{{\mathbf r}_i} + \Phi_{{\mathbf r}_i+{\mathbf e}_1} + \Phi_{{\mathbf r}_i+{\mathbf e}_2} + \Phi_{{\mathbf r}_i+{\mathbf e}_3)}.
\end{cases}
\label{eq:Gausslaw}
\end{eqnarray}
The different sign ($\pm$) reflects the fact that the electric charge is defined as a divergence of the electric fields:
$\Phi_{i\alpha}$ is defined as the field outward from the downward dual tetrahedron.

The Gauss' law, Eq.~(\ref{eq:Gausslaw}) is not sufficient to determine the electric fields in a unique way.
We further impose the rotation-free condition for the electric fields, $\{\Phi_{i\alpha}\}$.
Each site ${\mathbf R}_{i\alpha}$ is located at the center of a hexagonal ring of the dual pyrochlore lattice, and is surrounded by 6 sites on the dual lattice.
We impose the staggered summation of electric fields on these sites to be zero.
This condition can be expressed simply as
\begin{eqnarray}
\sum_{\beta, \gamma\not=\alpha, \beta\not=\gamma}(\Phi_{{\mathbf R}_{i\alpha} + ({\mathbf e}_{\beta} - {\mathbf e}_{\gamma})} - \Phi_{{\mathbf R}_{i\alpha} - ({\mathbf e}_{\beta} - {\mathbf e}_{\gamma})}) = 0,
\label{eq:Rotationfree}
\end{eqnarray}
for each site $(i, \alpha)$. Eqs.~(\ref{eq:Gausslaw}) and (\ref{eq:Rotationfree}) lead to the unique solution of the electric fields, $\{\Phi_{i\alpha}\}$, which satisfies the Poisson's equation.

Before showing the concrete form of the solution, we introduce the relation to $\{\theta_j\}$. The electric fields are connected to $\{\theta_j\}$ as their rotations.
\begin{widetext}
\begin{eqnarray}
\begin{cases}
\Phi_{{\mathbf r}_{i0}} + 2\pi W_{i0} = -(\theta_{{\mathbf r}_{i0} + ({\mathbf e}_1-{\mathbf e}_2)} - \theta_{{\mathbf r}_{i0} - ({\mathbf e}_2-{\mathbf e}_3)} + \theta_{{\mathbf r}_{i0} + ({\mathbf e}_3-{\mathbf e}_1)} - \theta_{{\mathbf r}_{i0} - ({\mathbf e}_1-{\mathbf e}_2)} + \theta_{{\mathbf r}_{i0} + ({\mathbf e}_2-{\mathbf e}_3)} - \theta_{{\mathbf r}_{i0} - ({\mathbf e}_3-{\mathbf e}_1)})\\
\Phi_{{\mathbf r}_{i1}} + 2\pi W_{i1} = \theta_{{\mathbf r}_{i1} + ({\mathbf e}_2-{\mathbf e}_3)} - \theta_{{\mathbf r}_{i1} - ({\mathbf e}_3-{\mathbf e}_0)} + \theta_{{\mathbf r}_{i1} + ({\mathbf e}_0-{\mathbf e}_2)} - \theta_{{\mathbf r}_{i1} - ({\mathbf e}_2-{\mathbf e}_3)} + \theta_{{\mathbf r}_{i1} + ({\mathbf e}_3-{\mathbf e}_0)} - \theta_{{\mathbf r}_{i1} - ({\mathbf e}_0-{\mathbf e}_2)}\\
\Phi_{{\mathbf r}_{i2}} + 2\pi W_{i2} = \theta_{{\mathbf r}_{i2} + ({\mathbf e}_3-{\mathbf e}_1)} - \theta_{{\mathbf r}_{i2} - ({\mathbf e}_1-{\mathbf e}_0)} + \theta_{{\mathbf r}_{i2} + ({\mathbf e}_0-{\mathbf e}_3)} - \theta_{{\mathbf r}_{i2} - ({\mathbf e}_3-{\mathbf e}_1)} + \theta_{{\mathbf r}_{i2} + ({\mathbf e}_1-{\mathbf e}_0)} - \theta_{{\mathbf r}_{i2} - ({\mathbf e}_0-{\mathbf e}_3)}\\
\Phi_{{\mathbf r}_{i3}} + 2\pi W_{i3} = \theta_{{\mathbf r}_{i3} + ({\mathbf e}_1-{\mathbf e}_2)} - \theta_{{\mathbf r}_{i3} - ({\mathbf e}_2-{\mathbf e}_0)} + \theta_{{\mathbf r}_{i3} + ({\mathbf e}_0-{\mathbf e}_1)} - \theta_{{\mathbf r}_{i3} - ({\mathbf e}_1-{\mathbf e}_2)} + \theta_{{\mathbf r}_{i3} + ({\mathbf e}_2-{\mathbf e}_0)} - \theta_{{\mathbf r}_{i3} - ({\mathbf e}_0-{\mathbf e}_1)}
\end{cases}
\label{eq:DefinitionElectricfield}
\end{eqnarray}
\end{widetext}
Here, each $\theta_j$ is defined in a range, $-\pi<\theta_j\leq\pi$, and the electric fields are also defined in the same range, $-\pi<\Phi_{i\alpha}\leq\pi$.
If the summation of $\theta_j$'s on the righthand side of the equation goes out of this range, it is compensated by the integers, $W_{i\alpha}$, so that $\Phi_{i\alpha}$ is within the range.
In Eq.~(\ref{eq:DefinitionElectricfield}), $\theta_j$ at the same site $j$ appears twice in the opposite sign.
It guarantees that the summation of electric fields to be integer multiple of $2\pi$. leading to the quantization of electric charge, according to Eq.~(\ref{eq:Gausslaw}).

To come back to finding the solution of $\Phi_{i\alpha}$, we first Fourier transform Eqs.~(\ref{eq:Gausslaw}) and (\ref{eq:Rotationfree}), and obtain
\begin{eqnarray}
\begin{cases}
Q_{{\mathbf q}d}=\frac{1}{2}(\Phi_{{\mathbf q}0} + \Phi_{{\mathbf q}1}e^{-i\frac{q_1}{2}} + \Phi_{{\mathbf q}2}e^{-i\frac{q_2}{2}} + \Phi_{{\mathbf q}3}e^{-i\frac{q_3}{2}}),\\
Q_{{\mathbf q}u}=-\frac{1}{2}(\Phi_{{\mathbf q}0} + \Phi_{{\mathbf q}1}e^{i\frac{q_1}{2}} + \Phi_{{\mathbf q}2}e^{i\frac{q_2}{2}} + \Phi_{{\mathbf q}3}e^{i\frac{q_3}{2}}),
\end{cases}
\label{eq:Gausslaw_Fourier}
\end{eqnarray}
and
\begin{eqnarray}
\begin{cases}
\Phi_{{\mathbf q},1}\frac{q_2-q_3}{2} + \Phi_{{\mathbf q},2}\sin\frac{q_3-q_1}{2} + \Phi_{{\mathbf q},3}\sin\frac{q_1-q_2}{2} = 0,\\
\Phi_{{\mathbf q},0}\sin\frac{q_2-q_3}{2} + \Phi_{{\mathbf q},2}\sin\frac{q_3}{2} - \Phi_{{\mathbf q},3}\sin\frac{q_2}{2} = 0,\\
\Phi_{{\mathbf q},0}\sin\frac{q_3-q_1}{2} + \Phi_{{\mathbf q},3}\sin\frac{q_1}{2} - \Phi_{{\mathbf q},1}\sin\frac{q_3}{2} = 0,\\
\Phi_{{\mathbf q},0}\sin\frac{q_1-q_2}{2} + \Phi_{{\mathbf q},1}\sin\frac{q_2}{2} - \Phi_{{\mathbf q},2}\sin\frac{q_1}{2} = 0.
\end{cases}
\label{eq:Rotationfree_Fourier}
\end{eqnarray}
Here, we have introduced the momentum representations of the electric charges and fields,
\begin{eqnarray}
\begin{cases}
Q_{{\mathbf q}d(u)} = \frac{1}{2}\sum_iQ_{id(u)}e^{-i{\mathbf q}\cdot{\mathbf r}_{i}},\\
\Phi_{{\mathbf q}\alpha} = \sum_i\Phi_{i\alpha}e^{-i{\mathbf q}\cdot{\mathbf r}_{i\alpha}},
\end{cases}
\label{eq:Gauss}
\end{eqnarray}
and momenta, $q_{\alpha}={\mathbf q}\cdot{\mathbf E}_{\alpha}$.

These sets equations can be rewritten concisely by introducing the vectors,
\begin{eqnarray}
\vec{\Phi}_{{\mathbf q}} = \begin{bmatrix}
\Phi_{{\mathbf q},1}\\
\Phi_{{\mathbf q},2}\\
\Phi_{{\mathbf q},3}
\end{bmatrix},\ \ 
\vec{s}_{\mathbf q} = \begin{bmatrix}
\sin\frac{q_1}{2}\\
\sin\frac{q_2}{2}\\
\sin\frac{q_3}{2}
\end{bmatrix},\ \ 
\vec{c}_{\mathbf q} = \begin{bmatrix}
\cos\frac{q_1}{2}\\
\cos\frac{q_2}{2}\\
\cos\frac{q_3}{2}
\end{bmatrix}, 
\end{eqnarray}
and the orthogonal coordinates, spanned by $\vec{s}_{\mathbf q}$, $\vec{s}_{\mathbf q}\times\vec{c}_{\mathbf q}$, and $\vec{s}_{\mathbf q}\times(\vec{s}_{\mathbf q}\times\vec{c}_{\mathbf q})$.
With these notations, Eqs.~(\ref{eq:Gausslaw_Fourier}) and (\ref{eq:Rotationfree_Fourier}) are rewritten as
\begin{eqnarray}
\begin{cases}
\Phi_{{\mathbf q},0} + \vec{\Phi}_{{\mathbf q}}\cdot\vec{c}_{\mathbf q} = Q_{{\mathbf q},a},\\
\vec{\Phi}_{{\mathbf q}}\cdot\vec{s}_{\mathbf q} = iQ_{{\mathbf q},s} ,\\
\Phi_{{\mathbf q},0}(\vec{s}_{\mathbf q}\times\vec{c}_{\mathbf q}) + \vec{\Phi}_{{\mathbf q}}\times\vec{s}_{\mathbf q} = 0,\\
\vec{\Phi}_{{\mathbf q}}\cdot(\vec{s}_{\mathbf q}\times\vec{c}_{\mathbf q}) = 0,
\end{cases}
\label{eq:Poisson_Fourier}
\end{eqnarray}
where we have introduced $Q_{{\mathbf q},s} = Q_{{\mathbf q},d} + Q_{{\mathbf q},u}$, $Q_{{\mathbf q},a} = Q_{{\mathbf q},d} - Q_{{\mathbf q},u}$.
By expanding $\vec{\Phi}_{{\mathbf q}} = \Phi_s\vec{s}_{\mathbf q} + \Phi_{sc}\vec{s}_{\mathbf q}\times\vec{c}_{\mathbf q} + \Phi_{ssc}\vec{s}_{\mathbf q}\times(\vec{s}_{\mathbf q}\times\vec{c}_{\mathbf q})$, and inputting it into Eq.~(\ref{eq:Poisson_Fourier}), we obtain the final expression,
\begin{widetext}
\begin{eqnarray}
\Phi_{{\mathbf q},0}=\frac{|\vec{s}_{\mathbf q}|^2Q_{{\mathbf q},a} - i(\vec{s}_{\mathbf q}\cdot\vec{c}_{\mathbf q})Q_{{\mathbf q},s}}{|\vec{s}_{\mathbf q}|^2 + |\vec{s}_{\mathbf q}\times\vec{c}_{\mathbf q}|^2},
\label{eq:Phi0_appendix}
\end{eqnarray}
\begin{align}
\vec{\Phi}_{{\mathbf q}} = \frac{1}{|\vec{s}_{\mathbf q}|^2 + |\vec{s}_{\mathbf q}\times\vec{c}_{\mathbf q}|^2}\Bigl[\vec{s}_{\mathbf q}iQ_{{\mathbf q},s} + |\vec{s}_{\mathbf q}|^2\Bigl(\vec{c}_{\mathbf q} - \frac{\vec{s}_{\mathbf q}}{|\vec{s}_{\mathbf q}|}\Bigl(\vec{c}_{\mathbf q}\cdot\frac{\vec{s}_{\mathbf q}}{|\vec{s}_{\mathbf q}|}\Bigr)\Bigr)Q_{{\mathbf q},a} + |\vec{c}_{\mathbf q}|^2\Bigl(\vec{s}_{\mathbf q} - \frac{\vec{c}_{\mathbf q}}{|\vec{c}_{\mathbf q}|}\Bigl(\vec{s}_{\mathbf q}\cdot\frac{\vec{c}_{\mathbf q}}{|\vec{c}_{\mathbf q}|}\Bigr)\Bigr)iQ_{{\mathbf q},s}\Bigr],
\label{eq:Phi_appendix}
\end{align}
\end{widetext}
which are referred to in the main text as Eqs.~(\ref{eq:Phi0_appendix}) and (\ref{eq:Phi_appendix}).

\section{Continuum limit}
\label{sec:CoulombInteraction}
In this section, we show that the electric fields approach the form expected in the classical electromagnetism in the continuum limit on the basis of the variational ansatz, and thereby derive the $1/r$-type interaction potential between electric monopoles.
To this aim, we start from the variational form of the electric fields, Eqs.~(\ref{eq:Phi0_appendix}) and (\ref{eq:Phi_appendix}), and show that it decays as $1/r^2$ from the position of fixed electric charge in the continuum limit.

We assume single charge $Q$ is located at the downward dual tetrahedron at the origin, $(0, 0, 0)$, then the electric charge takes the form,
\begin{eqnarray}
Q_{{\mathbf q}d}=\frac{Q}{2},\ \ \  Q_{{\mathbf q}u}=0.
\end{eqnarray}
Putting it into Eqs.~(\ref{eq:Phi0_appendix}) and (\ref{eq:Phi_appendix}), we obtain
\begin{eqnarray}
\Phi_{{\mathbf q}\alpha}=\frac{Q}{4}e^{-i\frac{q_{\alpha}}{2}}\frac{\sum_{\beta\not=\alpha}(e^{iq_{\alpha}} - e^{iq_{\beta}})}{|\vec{s}_{\mathbf q}|^2 + |\vec{s}_{\mathbf q}\times\vec{c}_{\mathbf q}|^2}.
\end{eqnarray}
Then, making Fourier transformation back to the real space, we obtain
\begin{align}
\Phi_{i\alpha}&=\frac{Q}{4}\frac{1}{N_{\rm c}}\sum_{\mathbf q}e^{i{\mathbf q}\cdot{\mathbf r_i}}\frac{\sum_{\beta\not=\alpha}(1 - e^{i(q_{\beta}-q_{\alpha})})}{|\vec{s}_{\mathbf q}|^2 + |\vec{s}_{\mathbf q}\times\vec{c}_{\mathbf q}|^2}\nonumber\\
&=\frac{1}{4}\sum_{\beta\not=\alpha}(\Psi({\mathbf r}_i) - \Psi({\mathbf r}_i + {\mathbf E}_{\beta} - {\mathbf E}_{\alpha})),
\label{eq:Electricfield_by_Pointcharge}
\end{align}
where we have introduced
\begin{eqnarray}
\Psi({\mathbf r})\equiv\frac{1}{N_{\rm c}}\sum_{\mathbf q}\frac{Qe^{i{\mathbf q}\cdot{\mathbf r}}}{|\vec{s}_{\mathbf q}|^2 + |\vec{s}_{\mathbf q}\times\vec{c}_{\mathbf q}|^2}.
\end{eqnarray}
In the long-wavelength limit, ${\mathbf q}\to0$, its denominator converges as
\begin{align}
&|\vec{s}_{\mathbf q}|^2 + |\vec{s}_{\mathbf q}\times\vec{c}_{\mathbf q}|^2\nonumber\\
&=\sin^2\frac{q_1}{2} + \sin^2\frac{q_2}{2} + \sin^2\frac{q_3}{2}\nonumber\\
 &\hspace{1cm}+ \sin^2\frac{q_1-q_2}{2} + \sin^2\frac{q_1-q_3}{2} + \sin^2\frac{q_2-q_3}{2}\nonumber\\
&\to q_x^2+q_y^2+q_z^2 = |{\mathbf q}|^2.
\end{align}
Consequently, $\Psi({\mathbf r})$ decays as $\propto\frac{1}{|{\mathbf r}|}$ in the long distance, i.e. $\Psi({\mathbf r})$ means the electrostatic potential created by a point charge, $Q$. 

Given this electrostatic interpretation of $\Psi({\mathbf r})$, we consider the long distance behavior of $\Phi_{i\alpha}$ based on Eq.~(\ref{eq:Electricfield_by_Pointcharge}).
If we set ${\mathbf r}_{i\alpha}$ far enough from the origin, we can approximate
\begin{eqnarray}
\Psi({\mathbf r}_i) - \Psi({\mathbf r}_i + {\mathbf E}_{\beta} - {\mathbf E}_{\alpha})\simeq -({\mathbf E}_{\beta} - {\mathbf E}_{\alpha})\cdot\nabla\Psi({\mathbf r}_i).
\end{eqnarray}
Putting this expression into Eq.~(\ref{eq:Electricfield_by_Pointcharge}), we obtain
\begin{eqnarray}
\Phi_{i\alpha}\simeq -\Bigl(\frac{1}{4}\sum_{\beta\not=\alpha}({\mathbf E}_{\beta} - {\mathbf E}_{\alpha})\Bigr)\cdot\nabla\Psi({\mathbf r}_i).
\end{eqnarray}
Here, the vector,
\begin{eqnarray}
\frac{1}{4}\sum_{\beta\not=\alpha}({\mathbf E}_{\beta} - {\mathbf E}_{\alpha}),
\end{eqnarray}
connects from the center of upward dual tetrahedron to the center of downward dual tetrahedron through the vertex, ${\mathbf r}_{i\alpha}$.
It means $\Phi_{i\alpha}$ is equal to the negative gradient of the electrostatic potential, $\Psi({\mathbf r})$, at ${\mathbf r}_{i\alpha}$, and behaves as the electric field produced by a point charge, in the sense of classical electromagnetism.

\bibliographystyle{apsrev4-2}
\bibliography{Emonopole_dynamics}

%
%\end{thebibliography}

\end{document}